\documentclass[reprint, amsmath,amssymb,aps,longbibliography]{revtex4-2}

\usepackage[bottom]{footmisc}

\usepackage{kpfonts}
\usepackage[T1]{fontenc}
\usepackage[latin9]{inputenc}
\usepackage{mathrsfs}
\usepackage{bm}
\usepackage{amsmath}
\usepackage{amsthm}
\usepackage{amssymb}
\usepackage{stmaryrd}
\usepackage{mathtools}
\usepackage{changepage}

\usepackage{comment}
\usepackage{graphicx}
\usepackage{mathtools}
\usepackage{dcolumn}
\usepackage{bm}
\usepackage{xr}

\DeclarePairedDelimiter\abs{\lvert}{\rvert}%
\makeatletter
\let\oldabs\abs
\def\abs{\@ifstar{\oldabs}{\oldabs*}}

\DeclareFontShape{OMX}{cmex}{m}{b}{<-> cmexb10}{}
\SetSymbolFont{largesymbols}{bold}{OMX}{cmex}{m}{b}

\usepackage{stackengine}
\stackMath

\edef\ordinarycolon{\mathchar\the\mathcode`: }
\edef\ordinaryequals{\mathchar\the\mathcode`= }

\usepackage{breqn}

\AtBeginDocument{%
	\catcode`^=12
}

\allowdisplaybreaks

\let\cat@comma@active\@empty

\usepackage{color}

\usepackage{hyperref}
\usepackage[capitalize]{cleveref}

\newif\ifnotes
\notestrue

\usepackage{color}
\RequirePackage[dvipsnames,usenames]{xcolor}

\renewcommand{\mid}{\vert}

\makeatother

\newcommand{\ba}{\begin{eqnarray}}
	\newcommand{\ea}{\end{eqnarray}}

\newcommand{\eq}[1]{\begin{align}#1\end{align}}

\newcommand{\lr}[1]{\langle #1 \rangle}

\newcommand{\T}{{\mathcal{T}}}

\newcommand{\AAA}{{\mathcal{A}}}

\newcommand{\NN}{\mathcal{N}}

\newcommand{\Lu}[1]{\mathcal{L}(p_{x_{#1}}(0),p_{x_{#1}}(\tau))}

\newcommand{\RM}[3]{K^{#1}_{#2}(#3 t)}

\newcommand{\PF}[3]{\RM{#1}{#2}{#3} p_{#1}(t)}

\newcommand{\EPnew}[2]{\lr{\sigma^{#1}(#2)}}
\newcommand{\Anew}[3]{\lr{\AAA^{#1}_{#3}}_{#2}}
\newcommand{\EPRnew}[2]{\lr{\dot{\sigma}^{#1}(#2)}}
\newcommand{\EPRi}[3]{\lr{\dot{\zeta}^{#1}_{#3}(#2)}}
\newcommand{\EPi}[3]{\lr{\zeta^{#1}_{#3}(#2)}}

\newcommand{\PFiforwardverbose}{\PF{x'_i, x'_{-i}}{x_i, x'_{-i}}{i;}}
\newcommand{\PFireverseverbose}{\PF{x_i, x'_{-i}}{x'_i, x'_{-i}}{i;}}

\usepackage{color}
\definecolor{myblue}{rgb}{.8, .8, 1}

\usepackage{amsmath}
\usepackage{empheq}

\newlength\mytemplen
\newsavebox\mytempbox

\begin{document}
	
	\preprint{APS/123-QED}
	
	
	\title{Thermodynamic speed limits for co-evolving systems}%
	
	\author{Farita Tasnim}
	\email[]{farita@mit.edu, web: farita.me}
	\affiliation{Massachusetts Institute of Technology, Cambridge, MA, USA}
	
	\author{David H. Wolpert}
	\email[]{dhw@santafe.edu, web: davidwolpert.weebly.com}
	\affiliation{Santa Fe Institute, Santa Fe, NM, USA}
	\affiliation{Complexity Science Hub, Vienna, Austria}
	\affiliation{Arizona State University, Tempe, AZ, USA}
	
	\date{\today}
	
	\begin{abstract}
    Previously derived ``global'' thermodynamic speed limit theorems state that increasing the maximum speed with which a system can evolve between two given probability distributions over its states requires the system to produce more entropy in its evolution. 
    However, these theorems ignore that many systems are not monolithic, but instead comprise multiple subsystems that interact according to an (often sparse) network. 
    Indeed, most naturally-occurring and human-engineered systems of increasing complexity can be decomposed into sets of 
    co-evolving subsystems, where there exist \textit{a priori} constraints on the dynamics of each subsystem, restricting which other subsystems can affect its dynamics. 
    Here we derive three new SLTs that account for the thermodynamic effects of such constraints. 
    Our first new speed limit strengthens the global speed limit. 
    While our other two SLTs do not have this guarantee, in some situations they are even stronger than our first speed limit.
    Our results establish that a stochastically evolving system will, on average, produce more entropy in evolving between two distributions within a given time simply due to its comprising multiple, co-evolving subsystems. 
    We illustrate our results with numerical calculations involving a model of two cells sensing and storing information about their environment.

	\end{abstract}
	\maketitle
	
	

\noindent \textbf{\large{Introduction}}

$ $

	
	

	
	We can characterize a stochastic process by the minimum time it takes
	to evolve from one particular, specified probability distribution over its states into another 
	specified distribution. 
	To give just a few examples, such lower bounds usefully describe: 
	a cell that senses a change in its environment~\cite{hartich_barato_seifert_2016, ashida2021experimentalSLT} or synthesizes a protein~\cite{dutta2020protein-synthesis};
	a set of chemical species that changes their concentrations via
	a chemical reaction network~\cite{rao2018chemrxn, rao2016chemrxnPRX}; 
    a genome that evolves to include particular mutations~\cite{colegrave2002sex, mcdonald2016sex}; a digital device that completes a particular 
    computation~\cite{lloyd2000ultimate, markov2014limits};
	a network of neurons that store, process 
	and transmit information~\cite{laughlin1998metabolic, laughlin2001energy, laughlin2003communication} or (re)configures itself~\cite{bassett2011dynamic, bullmore2012economy};
	and an opinion network that evolves from a unimodal to a bimodal state~\cite{li2019opinionconsensus, siegenfeld2020negative}.
    
    Continuous-time Markov chains (CTMC's) can accurately serve as models of many of these stochastic processes. 
    With the new body of work sometimes called (classical) ``stochastic thermodynamics"~\cite{esposito2010threefaces, seifert2012stochthermomolecularmachines,  vdb2013stochastic, vdbesposito2015ensemble, ciliberto2017experiments, wolpert2019stofcomputation} developed in the last two decades, we have drastically improved our ability to analyze the thermodynamics of CTMCs. 
    For example, thermodynamic uncertainy relations (TURs) \cite{horowitz2017proofFT-TUR, horowitz2020thermodynamic} have shown that dissipated work constrains the fluctuations in any current flowing in a system at a steady state arbitrarily far from equilibrium. 
    Additionally, fluctuation theorems (FTs) \cite{seifert2012stochthermomolecularmachines,esposito2010threeDFTs, rao2018DFT} have shown that probability distributions over any functional of a stochastic trajectory obey symmetries that constrain their even and odd moments.
    
    In addition to these powerful new results, the last few years have brought about a quickly growing set of lower bounds on the time it takes a system evolving according to a CTMC
    to move from one given distribution over the system's states to another one~\cite{ito2018stochastic, ito2020stochastic, sekimoto1997complementarity, aurell2012refined, vo_vu_hasegawa_2020, vanvu2021geometrical, shiraishi_funo_saito_2018}. 
    These ``thermodynamic speed limit theorems" (SLTs) reveal a trade-off between the time for the evolution to take place, and the amount of work dissipated (and, thus, free energy consumed) during that evolution.
    
    As a canonical example, the SLT derived in~\cite{shiraishi_funo_saito_2018} lower bounds this time of evolution with a quantity proportional to the total variation distance between the initial and final distributions over system states; and inversely proportional to the total entropy production (EP) and time-averaged dynamical activity. 
    When local detailed balance holds, the EP equals the work a system must dissipate during a process, i.e., the degree to which the total entropy of the universe increases, or the expended free energy that cannot be recovered. 
	Therefore, intuitively, this SLT declares that if we modify a CTMC to make it change its distribution either by a greater amount or in less time, then we must ``pay for'' that faster evolution by increasing (a lower bound on) the total dissipated work.
	
However, like most of the other theorems in (classical) stochastic thermodynamics~\cite{esposito2010threefaces, vdb2013stochastic, vdbesposito2015ensemble, pietzonka2017finitetimeTUR}, the previously derived SLTs do not
exploit any information concerning the internal structure of a system.
	In particular, these SLTs ignore all aspects of how the system might decompose into a set of co-evolving subsystems. 
	Although this attribute allows for their broad applicability, these SLTs will, in general, provide increasingly weak bounds for systems of increasing complexity. 
	As a result, such ``global SLTs'', along with other previously derived thermodynamic relations (such as the TURs and FTs) that formalize systems as monolithic entities, most appropriately apply to nanoscale
    systems~\cite{seifert2012stochthermomolecularmachines, collin2005verification, xiong2018experimental, huber2008employing, toyabe2010experimental, barato2015thermodynamic, friedman2020thermodynamic, pietzonka2016universal}, 
    whose internal structure is either unimportant (because the system is so small) or unknown. 
	
	Importantly, this shortcoming needlessly limits the strength of these results. 
	For many systems above the nanoscale, we know much about how their internal subsystems influence each other. 
	For example, many sets of co-evolving subsystems form modular interaction networks~\cite{dsouza2009structure, horowitz2014bipartiteinfoflow, horowitz2015multipartiteinfoflow, ito2013information, wolpert_bn_2020, wolpertstrengthenedsecondlaw, wolpert_fts_mpps_2020, wolpert_min_ep_2020, pilosof2017multilayer, delmas2019analysing, bassett2006small}. 
	This kind of internal structure can be formulated as a set of constraints on the CTMC governing the overall
	dynamics of the whole system.
	
	
	
	Here we derive tighter SLTs by analyzing how constraints on the allowed dynamics in a system affect its stochastic thermodynamics. 
	Specifically, we explore a major class of dynamical constraints that arises in systems of increasing complexity, as imposed by the property that they decompose into sets of co-evolving subsystems, where the interactions between subsystems restrict how the overall, combined system can evolve. 

	\begin{figure}[b]
    	\includegraphics[width=0.48\textwidth]{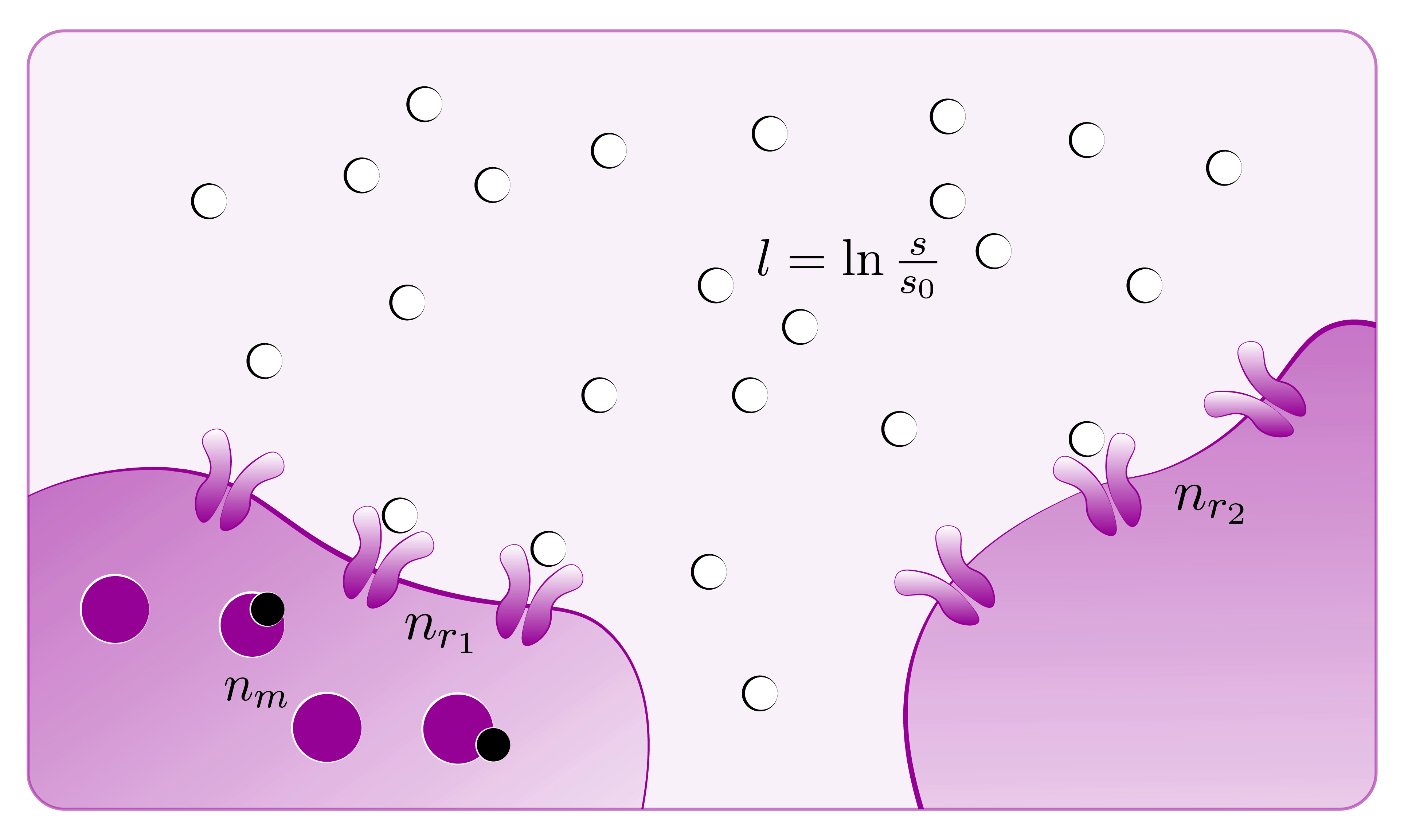}
    	\caption{
    	An example of a multipartite process. 
    	This MPP consists of four subsystems and maps to the abstracted MPP given in~\cref{fig:MPPExDef}. 
    	Subsystem 3 is $l$, the concentration of the ligand in a cellular medium. Two "nearby" cells sense this ligand concentration. The number of bound receptors, $n_{r_1}$, in Cell 1 comprise subsystem 2, and the number of bound receptors, $n_{r_2}$, in Cell 2 comprise subsystem 4. Subsystem 1 represents a memory of (phosphorylated) proteins, $n_{m}$, in Cell 1 that observes the state of its receptors.
    	} \label{fig:ex}
    \end{figure}

	
	We begin by reviewing the relevant stochastic thermodynamics of co-evolving subsystems, with emphasis on subsystem-indexed contributions to thermodynamic quantities such as the entropy production. 
	As commonly done in the literature, we assume that subsystems are spatially separated, and therefore that the overall system evolves as a multipartite process (MPP) \cite{horowitz2014bipartiteinfoflow, horowitz2015multipartiteinfoflow, wolpert_min_ep_2020, wolpert_fts_mpps_2020}. 
	
	We next derive three major results, each a new SLT applicable to multipartite processes. 
	The first SLT demonstrates that accounting for the dynamical constraints resulting from the internal structure of interactions in an MPP strengthens the bound on the minimum time required for that system to evolve between two distributions. 
	In deriving our second SLT, we note that an important concept in analyzing a set of co-evolving subsystems is that of a "unit", which is a subset of subsystems whose joint dynamics is independent of the state of all the other subsystems. 
	While subsystems outside of a unit can ``observe'' (depend on the state of) a subsystem in a unit, the reverse is not true. 
	This SLT shows that a system can never evolve faster than its slowest-evolving unit. 
	Our third SLT shows that a system can never evolve faster than its slowest-evolving subsystem. 
	
	We prove that the first SLT is always at least as strong as the global SLT in~\cite{shiraishi_funo_saito_2018}, regardless of the details of the system's dependency constraints. 
    The other two SLTs do not always have this guarantee; however, in many cases, they are stronger than our first SLT. 
    We derive our new SLTs in full detail in Appendices S1 and S2. 
	
    We then conduct numerical calculations to explore the strength of our new SLTs for the example MPP of a cell sensing its environment, a model previously studied in the literature~\cite{hartich_barato_seifert_2016, brittain2017learningrate}. 
    This example, depicted in~\cref{fig:ex}, consists of two sets of cellular receptors and a cellular memory that sense and record the value of a changing ligand concentration. 
	
	We end the main text with a discussion of the significance of our results, 
	a summary of thermodynamic consistency and auxiliary results (including new speed limits for Bayes' Nets) derived in the sections S3 - S10 of the Appendices, and suggestions for future work. 
	
	We summarize all notation used in the main text in Appendix S11. 
	
	\begin{figure}[t]
		\includegraphics[width=0.48\textwidth]{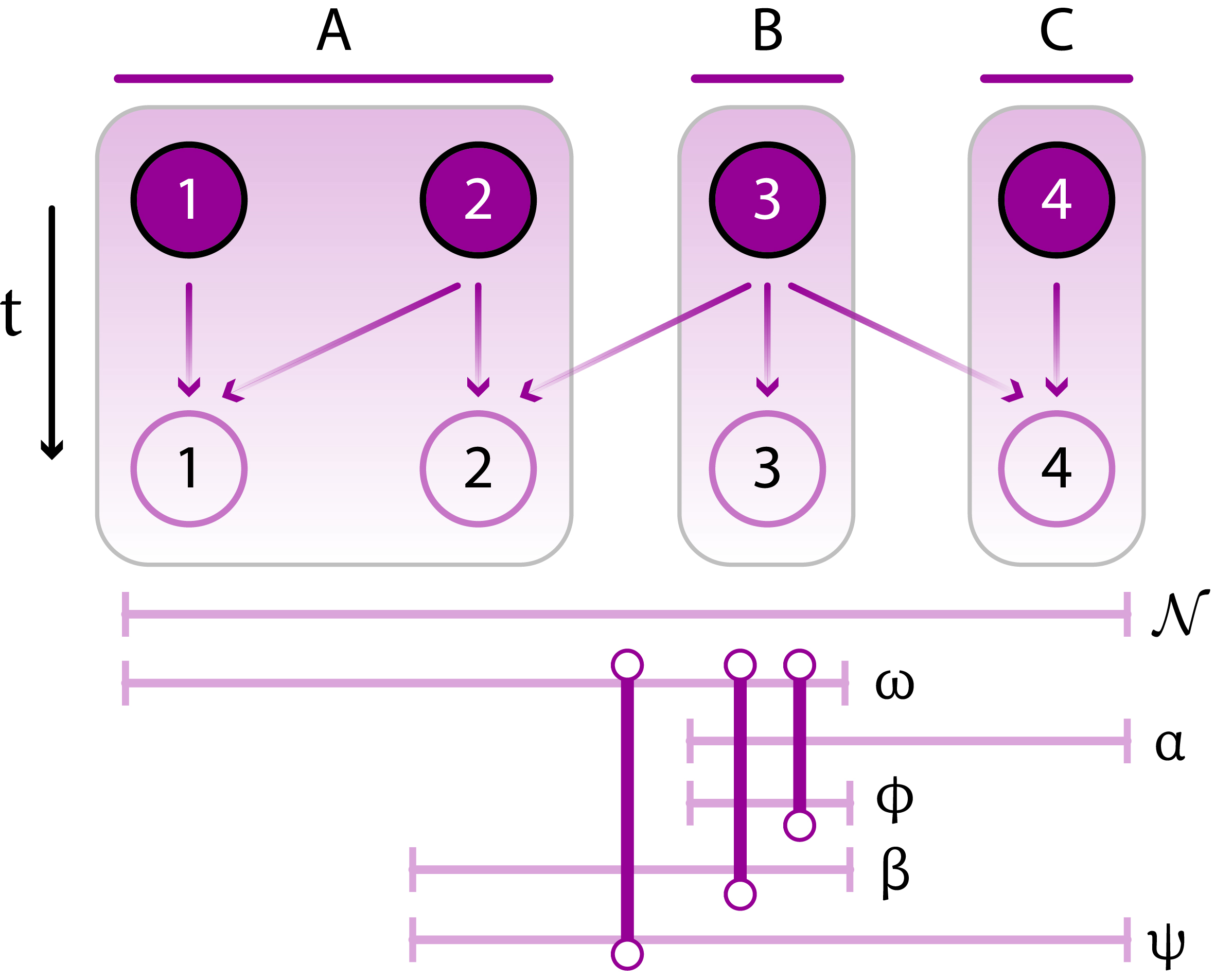}
		\caption{An example of a multipartite process. 
			The system consists of four subsystems, indicated by outlined circles. The arrows represent dependencies that restrict the associated subsystem's rate matrices:
			the state of subsystem 1 at time $t + \delta t$ depends on the state of subsystems 1 and 2 at time $t$; subsystem 2 depends on 2 and 3; subsystem 3 depends only on itself, and subsystem 4 depends on 3 and 4. 
			The set of all dependency constraints can be represented as a directed acyclic graph (DAG), called a \textbf{dependency graph} $ ( 1 \leftarrow 2 \leftarrow 3 \rightarrow 4 ) $. 
			Five units exist in the system aside from the global system itself: $\{ \omega, \alpha, \phi, \beta, \psi \}$. 
			Each\textbf{ unit structure}, indicated by circle-capped vertical lines, is a set of units that is closed under intersections and that covers $\NN$. 
			There are four possible unit structures, $\mathcal{N}^*_1 = \{ \omega, \alpha, \phi \}$, $\mathcal{N}^*_2 = \{ \omega, \alpha, \phi, \beta \}$, $\mathcal{N}^*_3 = \{ \omega, \psi, \beta \}$, or $\mathcal{N}^*_4 = \{ \omega, \psi, \alpha, \beta, \phi \}$.
		} \label{fig:MPPExDef}
	\end{figure}
	
	$ $
	$ $
	
\noindent \textbf{\large{Results}}

$ $
	
\noindent {\large{Stochastic thermodynamics of multipartite processes}}
	
	A set of co-evolving systems can be modeled to evolve as a multipartite process (MPP).
	Formally, an MPP describes any system with the property that no two of its subsystems can change state at \textit{exactly} the same time. 
	We call this the \textbf{multipartite property}. 
	MPPs are extremely common~\cite{horowitz2014bipartiteinfoflow, horowitz2015multipartiteinfoflow, wolpert_min_ep_2020, wolpert_fts_mpps_2020}, including biological sensors, information engines, and Ising spin systems that can be modeled with the Gillespie algorithm or kinetic Monte Carlo simulations~\cite{novinger2021particle}. 
	Other examples include eukaryotic cells, which consist of multiple interacting organelles and biomolecular species~\cite{hartich_barato_seifert_2016}. 
	Subcellular processes such as protein synthesis could also be modelled as MPPs, with subsystems as ribosomal subunits, mRNA, and sets of tRNA-activated amino acids~\cite{dutta2020protein-synthesis}. 
	An MPP can have any arbitrary collection of dependency constraints, i.e., restrictions on how the dynamics of each
	subsystem depends on the states of the other subsystems. 
	This includes non-reciprocal couplings between subsystems~\cite{loos2020irreversibility}, such as those that can be found in networks represented by directed acyclic graphs (DAGs)~\cite{eissfeller1994asymmetricising}. 
	This assumption of no ``back-action'' is used throughout literature~\cite{sagawa2009minimal, hartich_barato_seifert_2016}. 
	
	The ubiquity of MPPs arises because, unless two subsystems occupy the exact same location, they will not
	be coupled to the same degrees of freedom in the external reservoirs. 
	Therefore, it is physically impossible for them to change state at the \textit{exact} same moment due to a single fluctuation in their interaction with a reservoir. 
	Additionally, continuous time also allows for successive state transitions to occur arbitrarily closely in time.
	
	We consider an MPP $\mathcal{N}$ that comprises a set of $N$ subsystems with finite state spaces $\{ X_i : i = 1, \ldots , N \}$. $x$ indicates a state vector in $X =
	\bigtimes_{\;i \in \mathcal{N}} X_i $, the joint state space of the full system. 
	For any set of subsystems $A \subset \mathcal{N}$, we write $-A := \mathcal{N} \backslash A$,
	and $x_A$ indicates an element of $X_A = \bigtimes_{\;i \in A} X_i$. 
	For any two distributions $p$ and $p'$ over $X$, we write the total variation (L1) distance between them as
	\eq{
	    \mathcal{L}(p, p') = \sum_{x} \abs{p_{x} - p'_{x}}
	}
	We indicate the Shannon entropy of a distribution over states $X_A$ as $S^{X_A}$ or $S^A$. 
	We set $k_{\text{B}} = 1$. 

	The joint dynamics over $X$ is given by the master equation
	\eq{ \label{global.master.equation}
		\frac{dp_x(t)}{dt} = \sum_{x'} K^{x'}_{x}(t) p_{x'}(t)
	} 
	In an MPP, there also exists a set of $N$ (possibly time-varying) stochastic rate matrices, 
	$\{ K^{x'}_{x} (i; t) : i = 1, \ldots , N \}$
	where $K^{x'}_{x} (i; t) = 0  \; \forall \, i, t, \{ x', x \, \mid \, x'_{-i}  \neq x_{-i} \}$, 
	and we can decompose 
	\eq{
        K^{x'}_x(t) = \sum_{i \in \mathcal{N}} K^{x'}_{x}(i; t) 
	}
    Similarly, for any $B \subseteq \mathcal{N}$, $K^{x'}_{x} (B; t) = \sum_{i \in B} K^{x'}_{x} (i; t)$. 
    
    In an MPP, $K^{x'}_x(t) = 0$ if $x'$ and $x$ differ in more than
	one component. In other words, at any given moment in time, only one subsystem can undergo a state transition. This means that the global rate matrix is sparse for many MPPs.

	Note that for all $x' \neq x$,
	\eq{ \label{eq:dpx-dt}
		\RM{x'}{x}{i;} 
		= \RM{x'}{x}{} \delta (x'_{-i}, x_{-i})
	}
	For each subsystem $i$, we write $r(i; t)$ for any set of subsystems at time $t$ that include $i$ and for which we can write
	\eq{ \label{eq:global-rm-construction-from-leader-rms}
		K^{x'}_{x} (i; t) = K^{x'_{r(i;t)}}_{x_{r(i;t)}} (i; t) \, \delta (x'_{-r(i;t)},x_{-r(i;t)})
	} 
	In general, for any given $i$ there could be multiple such sets $r(i;t)$. 
	We refer to a specification of any $r(i; t)$ as a \textbf{dependency constraint}.

	We define a \textbf{unit} $\omega$ (at an implicit time $t$) as a set of subsystems $i$ such that $i \in \omega$ implies that $r(i;t) \subseteq \omega$. Consequently, any intersection of two units is a unit, as is any union of two units. 
	For any unit $\omega$, we can decompose 
    $K^{x'_{\omega}}_{x_{\omega}} (\omega; t) = \sum_{i \in \omega} K^{x'_{\omega}}_{x_{\omega}} (i; t)$. So, by~\cref{eq:global-rm-construction-from-leader-rms},  $K^{x'}_{x} (\omega; t) = \sum_{i \in \omega} K^{x'}_{x} (i; t) = K^{x'_{\omega}}_{x_{\omega}} (\omega; t) \delta (x'_{-\omega},x_{-\omega})$.

	We write $\NN^{**}$ for the set of all units in $\NN$, other than the global system itself. 
	For the example system shown in~\cref{fig:MPPExDef}, $\NN^{**} = \{ \omega, \alpha, \phi, \beta, \psi \}$. 
	For later use, we also define $\NN^\dagger = \NN^{**} \cup \NN$. 
	
	As is standard in the recent literature on MPPs~\cite{wolpert_fts_mpps_2020, wolpert_min_ep_2020, wolpertstrengthenedsecondlaw}, we assume that there are pre-fixed time intervals in which $\NN^{**}$ doesn't change, and restrict attention to such an interval.

    Crucially, the local marginal distribution $p_{x_{\omega}} (t)$ of any unit $\omega$ at any time $t$ evolves as a self-contained CTMC with the local rate matrix $K^{x'_{\omega}}_{x_{\omega}} (\omega; t)$ (proved in Appendix A of~\cite{wolpert_min_ep_2020}): 
	\eq{ \label{local.master.equation}
		\frac{d p_{x_{\omega}} (t)}{dt} = \sum_{x'_{\omega}} K^{x'_{\omega}}_{x_{\omega}} (\omega; t) p_{x'_{\omega}} (t) = \sum_{x'_{\omega}} \sum_{i \in \omega} K^{x'_{\omega}}_{x_{\omega}} (i; t) p_{x'_{\omega}} (t)
	}
    This means that any unit obeys all the usual stochastic thermodynamic theorems for CTMCs, e.g. the second law, FTs, TURs, and SLTs. 
    In general, this property does not hold for any single subsystem or any set of subsystems (that is not a unit) in an MPP, due to each subsystem's dependency constraints~\cite{wolpert_min_ep_2020}. 
    Instead, the marginal distribution of subsystem $i$ changes as
	\eq{
		&\frac{d}{dt} p_{x_i}(t) = \sum_{x'} \RM{x'}{x}{i;} p_{x'}(t) \label{eq:dpxi-dt} \\
		&= \sum_{x' \neq x} \RM{x'}{x}{i;} p_{x'}(t) - \RM{x}{x'}{i;} p_{x}(t)\\
		&= \sum_{x'_{-i}} \sum_{x'_i \neq x_i} \PFiforwardverbose - \PFireverseverbose
	}
	For any MPP, we write the global time-averaged dynamical activity (often interpreted as the frequency of state transitions) during the time interval $[0, \tau]$ as 
	\eq{
	    \Anew{\NN}{\tau}{} = \frac{1}{\tau} \int_0^\tau \sum_{x' \neq x} dt \; K^{x'}_{x}(t)p_{x'}(t)
	}
    We write the global EP rate as
	\eq{
	    \EPRnew{\NN}{t} = \sum_{x', x} K^{x'}_{x}(t)p_{x'}(t) \ln \left[ \frac{K^{x'}_{x}(t)p_{x'}(t)}{K^{x}_{x'}(t)p_{x}(t)} \right]
	\label{eq:global_EP_def}
	}
	Integrating the global EP rate over the interval $[0, \tau]$ provides the total global EP $\EPnew{\NN}{\tau}$. 
	
	In our notation, the major result of~\cite{shiraishi_funo_saito_2018} --- ``global SLT'' --- reads $ \tau \geq \T := \frac{\left( \Lu{} \right)^2}{2 \EPnew{\NN}{\tau} \Anew{\NN}{\tau}{}}$. 
	Note that although the global SLT can be applied to an MPP, it does not account for any of the MPP's dependency constraints.

	In order to derive SLTs that account for the multipartite nature of the system dynamics, we decompose the global EP rate $\EPRnew{\NN}{t}$ for an MPP at time $t$:
    \eq{
		&\EPRnew{\NN}{t} = \sum_{i \in \NN} \sum_{x', x} \PF{x'}{x}{i;} \ln \left[ \frac{\PF{x'}{x}{i;}}{\PF{x}{x'}{i;}} \right] \label{eq:subsystem-rm-log} \\
		&= \sum_{i \in \NN} \sum_{x'_{-i}, x'_i, x_{i}} \PF{x'_i, x'_{-i}}{x_i, x'_{-i}}{i;} \ln \left[ \frac{\PF{x'_i, x'_{-i}}{x_i, x'_{-i}}{i;}}{\PF{x_i, x'_{-i}}{x'_i, x'_{-i}}{i;}} \right]  \\
		&:= \sum_{i \in \NN} \EPRi{i}{t}{\NN}
	}
	$\EPRi{i}{t}{\NN}$ is a subsystem-indexed contribution to the global EP rate, i.e., it changes only due to the state transitions in subsystem $i$ (compare it to the integrand in~\cref{eq:global_EP_def}). 
	We make the associated time-integrated definition, $\EPi{i}{\tau}{\NN} = \int_0^\tau dt \EPRi{i}{t}{\NN}$; therefore, $\EPnew{\NN}{t} = \sum_{i \in \NN} \EPi{i}{t}{\NN}$ for all times $t$. 
	Similarly, for any unit $\omega$,  we can write $\EPRnew{\omega}{t} := \sum_{i \in \omega} \EPRi{i}{t}{\omega}$ and $\EPnew{\omega}{t} = \sum_{i \in \omega} \EPi{i}{t}{\omega}$. 
	We define the vector whose components are the time-$t$ subsystem-indexed contributions to the global EP as $\lr{\vec{\zeta}_{\NN}(t)}$, and to the $\omega$-local EP as $\lr{\vec{\zeta}_{\omega}(t)}$. 
	
	Similarly, we define the dynamical activity due only to the state transitions in subsystem $i$ as:
	\eq{
		\AAA^i(t) := \sum_{x'_{-i}} \sum_{x_i} \sum_{x'_i \neq x_i} \PF{x'_i, x'_{-i}}{x_i, x'_{-i}}{i;}
	}
	where  $\AAA^{\NN}(t) := \sum_{x' \neq x} \PF{x'}{x}{} = \sum_{i \in \NN} \AAA^{i}(t)$. 
	The corresponding time-averaged dynamical activity 
    is $\Anew{i}{\tau}{} = \frac{1}{\tau} \int_0^\tau \AAA^{i}(t) $. 
	Similarly, for any unit $\omega$,  $\AAA^{\omega}(t) = \sum_{i \in \omega} \AAA^{i}(t)$ and $\Anew{\omega}{\tau}{} = \sum_{i \in \omega} \Anew{i}{\tau}{}$. 
	Unlike its entropy production, a subsystem's contribution to the dynamical activity does not depend on whether that dynamical activity is a global or local quantity (see Appendix S3 for proof). 
	We write $\lr{\vec{\AAA}_{\NN}}_t$ for the vector with components $\AAA^i(t)$, and write $\lr{\vec{\AAA}_{\omega}}_t$
	for the vector with components $\AAA^{\omega}(t)$.
	
	We formulate our analysis below as if each subsystem $i$ interacts wih a single heat bath (of inverse temperature $\beta_i$); however, all of our results extend naturally to the case of multiple reservoirs per subsystem, as discussed in~\cite{wolpert_fts_mpps_2020}.
	As a final comment, we emphasize that all thermodynamic speed limits hold for any chosen run-time $\tau$. 
	Indeed, since all EPs and dynamical activities depend on $\tau$, the speed limit bounds are functions of $\tau$.

$ $
$ $

\noindent {\large{Strengthened speed limits for multipartite processes}}

    Our first main result, derived in Appendix S1, states 
	\eq{ \label{eq:mpp-SLT-sub-N}
		\tau \geq \T_\NN := \frac{\left( \Lu{} \right)^2}{2 \left( \sum_{i \in \NN} \sqrt{\EPi{i}{\tau}{\NN} \Anew{i}{\tau}{}} \right)^2}
	}
	Furthermore, the Cauchy-Schwartz inequality ensures that $\left( \sum_{i \in \NN} \sqrt{\EPi{i}{\tau}{\NN} \Anew{i}{\tau}{}} \right)^2 \leq \EPnew{\NN}{\tau} \Anew{\NN}{\tau}{}$. Combining establishes that
	$\T_\NN \ge \frac{\left( \Lu{} \right)^2}{2 \EPnew{\NN}{\tau} \Anew{\NN}{\tau}{}} = \T$.
	
%
	Therefore, accounting for a system's internal structure strengthens the thermodynamic speed limit. 
	
	Our second main result is a speed limit involving only the thermodynamic contributions (and thus the dependency constraints) for the subsystems within any single unit $\omega \subset \NN$ (proof in Appendix S1):
	\eq{ \label{eq:mpp-SLT-sub-unit}
		\tau \geq \max_{\omega \in \NN^{**}} \left(
		\T_\omega :=\frac{\left( \Lu{\omega} \right)^2}{2 \left( \sum_{i \in \omega} \sqrt{\EPi{i}{\tau}{\omega} \Anew{i}{\tau}{}} \right)^2} \right)
	}
	In this sense, the slowest of all the units is the limiting factor on how fast the overall system can evolve.
	Below we will refer to the set of SLTs based on all of the $\T_\omega$ as the \textbf{unit-based SLTs}. 
	
	Our third main result bounds an MPP's speed of evolution using only the thermodynamic contributions of any single subsystem (proof in Appendix S2): 
	\eq{ \label{eq:mpp-SLT-sub-only}
		\tau \geq \max_{i \in \NN } \left( \T_i := \frac{\left( \Lu{i} \right)^2}{\min_{\omega \in \NN^\dagger \, | \, i \in \omega} 2 \EPi{i}{\tau}{\omega} \Anew{i}{\tau}{}} \right)
	} 
	In general, the unit satisfying the minimum in the denominator will be given by the smallest unit containing $i$. 
	Intuitively, this result shows that an MPP can evolve only as fast as its slowest-evolving subsystem. 
	The set of all $\T_i$ form the \textbf{subsystem-local SLTs}. 
	
	These second two results are not guaranteed to be tighter than the global SLT. However,
	in many cases, at least one of each kind of SLT (i.e.,~\cref{eq:mpp-SLT-sub-unit} for some unit $\omega$ or~\cref{eq:mpp-SLT-sub-only} some subsystem $i$) is stronger than even the bound given in our first result,~\cref{eq:mpp-SLT-sub-N}. 
    In general, which of our three main results will provide the tightest bound will vary depending on the details of the CTMC, particularly on the dependency graph and the control protocol (time sequence of rate matrices). 
	We can easily write necessary and sufficient conditions for any one of the three SLTs to be stronger than any one of the others. 
	For instance, a necessary and sufficient condition for~\cref{eq:mpp-SLT-sub-only} (for subsystem $i$) to outperform~\cref{eq:mpp-SLT-sub-N} is if the CTMC satisfies $\frac{\left( \sum_{j \in \NN} \sqrt{\EPi{j}{\tau}{\NN} \Anew{j}{\tau}{}} \right)^2}{\min_{\omega \in \NN^\dagger \, | \, i \in \omega} \EPi{i}{\tau}{\omega} \Anew{i}{\tau}{}} (\Lu{i})^2 \geq (\Lu{})^2$.  

    \begin{figure*}
    	\includegraphics[width=0.95\textwidth]{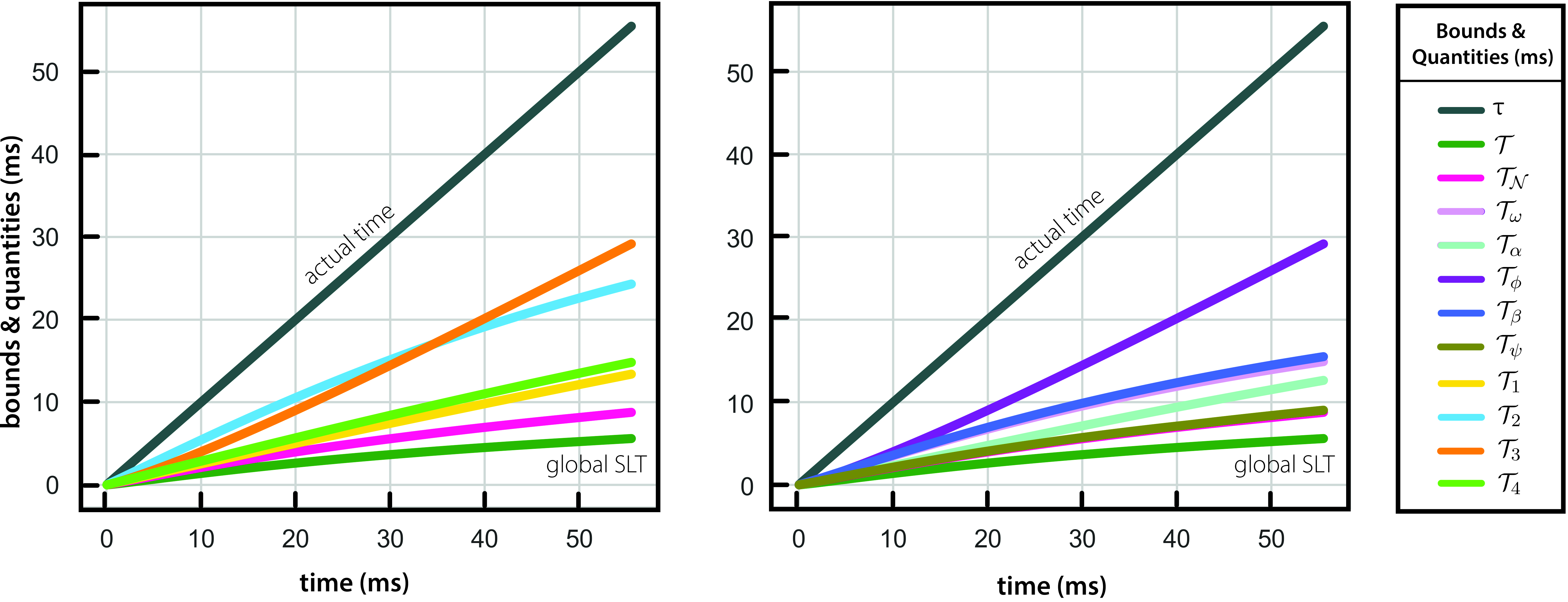}
    	\caption{Comparison of speed limit bounds for the example of cellular sensing shown in~\cref{fig:ex}. The dark grey line represents the actual time of system evolution. All other lines represent the lower bounds on time provided by different SLTs. Note that bounds represented by the orange and violet lines are equivalent because the unit $\phi$ is composed of only subsystem 3.} \label{fig:calcs}
    \end{figure*}

$ $
$ $

\noindent {\large{Numerical calculations of speed limit bounds for a model of cellular sensing}}

	We now illustrate our results for an MPP model of a cell sensing and storing information about its environment~\cite{hartich_barato_seifert_2016, brittain2017learningrate, wolpert_min_ep_2020}. This system is illustrated in~\cref{fig:ex} and captured abstractly in~\cref{fig:MPPExDef}. 
	Two sets of cellular receptors, belonging to different cells, each independently observe the concentration of a ligand in the medium. 
	Additionally, a set of proteins in one of the cells acts as a subcellular memory of that cell's fraction of bound receptors.

	Subsystem 3 is the ligand concentration, whose state is $l = \ln \frac{s}{s_0}$, where $s_0$ is a reference concentration. 
	Subsystem 2 is the number of receptors bound by the ligand in the cell membrane of Cell 1. 
	Its possible states are $n_{r_1} = 0, 1, 2, \ldots , N_{r_1}$, where $N_{r_1}$ is the total number of receptors. 
    Subsystem 4 is the number of bound receptors in Cell 2. 
	Its possible states are $n_{r_2} = 0, 1, 2, \ldots , N_{r_2}$, where $N_{r_2}$ is the total number of receptors. 
	Subsystem 1 is the number of phosphorylated proteins in Cell 1, serving as a memory of that cell's fraction of bound receptors. 
	Its possible states are $n_{m} = 0, 1, 2, \ldots , N_{m}$, where $N_{m}$ is the total number of proteins. 
    In our calculations, we use $N_{r_1} = 3, N_{r_2} = 3$, and $N_m = 4$.
	
	We construct the time-homogeneous rate matrices governing the evolution of this model system according to Section III of~\cite{hartich_barato_seifert_2016}. 
	These rate matrices account for quantities such as the free energy required for a ligand-receptor binding event and for a phosphorylation reaction.
	We set the initial joint distribution as $p_x(0) = p_{x_3}(0)p_{x_2 \mid x_3}(0)p_{x_4 \mid x_3}(0)p_{x_1 \mid x_2}(0)$, where each of the conditional distributions are Gaussians with mean set by the value of the state on which it is dependent, and where the initial distribution of subsystem 3, $p_{x_3}(0)$, is $\sim \NN(0, 0.01)$. 
	For more details, refer to the code available at  \url{https://github.com/FaritaTasnim/MPP_SLTs_cellular_sensing}. 
	
	We evolve the joint distribution over time according to the global rate matrix $K$ by solving the master equation to obtain $p_x(t) = p_x(0) e^{t K}$. We calculate the distribution every 50 $\mu$s in the interval [0, 55] ms. 
	From the rate matrices and time-$t$ distributions, we calculate all relevant thermodynamic quantities at each timestep.
	
	Our numerical calculations are illustrated in~\cref{fig:calcs}, for both subsystem-local SLTs (left panel) and for unit-based SLTs (right panel). 
	These calculations confirm that our first main result ($\T_\NN \equiv$ pink) provides a tighter bound on the speed of system evolution than does the global SLT from~\cite{shiraishi_funo_saito_2018} ($\T \equiv$ dark green).

	Additionally, we find that in this particular example, each unit-based SLT ($\T_\omega \equiv$ lavender, $\T_\alpha \equiv$ sea green, $\T_\phi \equiv$ violet, $\T_\beta \equiv$ dark blue, $\T_\psi \equiv$ olive green), as well as each subsystem-local SLT ($\T_1 \equiv$ yellow, $\T_2 \equiv$ light blue, $\T_3 \equiv$ orange, $\T_4 \equiv$ lime green), is stronger than not only the global SLT, but also our first SLT. 
	(As an aside, the units that minimize the denominator in the bound $\T_i$ for each subsystem $i$ are as follows: $\{ 1: \omega, \; 2: \beta, \; 3: \phi, \; 4: \alpha \}$.)
	As noted before, each of these units is the smallest unit containing $i$.
	
	Interestingly, even though the rate matrix for system evolution doesn't change (i.e., it is time-homogeneous) we see that the strength of the subsystem-local SLT for subsystem 3 ($\T_3 \equiv$ orange) surpasses that for subsystem 2 ($\T_2 \equiv$ light blue) after 35 ms of system evolution.

$ $

$ $

\noindent \textbf{\large{Discussion}}
	

	In this paper we extended the conventional thermodynamic speed limit to derive a set of strengthened speed limits for the case of co-evolving subsystems. These results can be useful for analyzing naturally occurring systems of many types
	where one would suppose there exist design pressures to make the evolution of an MPP as fast as possible, e.g.,
	in a biological setting, where speed might directly translate into fitness value.
	These results could additionally augment the evaluation of trade-offs between control protocols for evolving a system along a desired path of distributions~\cite{bravetti2017counteradiabatic, kolchinsky2020protocol-constraints, iram2021controllingcounteradiabatic}.

	We present other interesting properties related to SLTs in MPPs, as well as some auxiliary results in Appendices S4 - S10. 
	In Appendix S4, we discuss how our results extend the applicability of thermodynamic speed limits to systems characterized by local, rather than global, interactions. 
	In doing so, we establish thermodynamic consistency of our results for systems whose dynamics are defined by local, rather than global, Hamiltonians.
	In Appendix S5, we show that each of the subsystem-indexed contributions to the global EP follows an integral fluctuation theorem. 
	In Appendix S6, we note that the global SLT also applies to subsets of subsystems, or units, that evolve according to their own CTMC; this leads to what we call ``unit-local SLTs''. 
	In Appendix S7, we prove that the strongest of a generalized form of the SLT that involves a linear combination of unit-local thermodynamic properties is given by the unit-local SLT for the system's slowest-evolving unit. 
	In Appendix S8, we consider the SLTs as bounds on the space of distributions that a system can access within a given time. In particular, we show that the unit-local SLTs together more tightly restrict (than the global SLT) this space of final joint distributions for a pair of independently evolving spins. 
	In Appendix S9, we derive maximum and minimum speeds of evolution for Bayes' Nets, which are a type of MPP for which the state transitions occur according to a global clock. 
	In Appendix S10, we derive lower bounds on the difference between the EP rate and the rate at which one unit ``learns'' about another using counterfactual thermodynamic quantities. 
	
	Finally, we note that the results in this paper suggest several avenues for future work. 
	Our results investigate how the dynamical constraints imposed by the interaction network of co-evolving subsystems that comprise an MPP demand a decrease in the system's maximum possible speed of evolution. 
	However, we note that the effect on speed due to  other types of dynamical constraints should be explored. 
	In general, \textit{any} constraints on a system's CTMC will contribute
	to its minimal EP, strengthening the Second Law of thermodynamics. 
	As a practical matter, for any CTMC that obeys a given set of constraints while also implementing a given conditional distribution $P(x_f | x_i)$ on an initial distribution $P(x_i)$, often the EP is the dominant contributor to the total heat dissipation of the system. 
	This contribution of the EP to the thermodynamic cost of evolution often far exceeds the minimal cost established by the generalized Landauer bound, which is simply given by the change in the entropy of the system~\cite{diamantini2014generalizedlandauer, kempes2017thermodynamicefficiency}.
	
	Although dynamical constraints govern the thermodynamic costs of 
	classical, many-degree-of-freedom systems, the analysis of the thermodynamic implications of constraints on a system's allowed dynamics remains in its infancy. 
	One example of relevant research derives the stochastic thermodynamics under protocol constraints~\cite{kolchinsky2020protocol-constraints}. 
	There has also been some important work where the ``constraint'' on a many-degree-of-freedom classical system is simply that it be some very narrowly defined type of system, whose dynamics is specified by many different kinds of parameters. 
	For example, there has been analysis of the stochastic thermodynamics of chemical reaction networks~\cite{wachtel2018thermodynamically, rao2016chemrxnPRX, rao2018chemrxn, yoshimura2021thermodynamic}, of electronic circuits~\cite{gao2021circuits, wolpert2020circuits, freitas2020circuits}, of spin glasses where all spins are coupled to one another~\cite{herpich2020stochastic}, of biological copying mechanisms~\cite{poulton2019nonequilibrium}, and of systems in which the state transitions of subsystems occur according to a synchronous global clock~\cite{wolpert_bn_2020, ito2013information}. 
	
	The analyses in this paper as well as those in~\cite{wolpert_bn_2020, wolpert_fts_mpps_2020, wolpert_min_ep_2020, kardes2021turs} consider time-homogeneous dependency constraints. 
	In many real-world scenarios, however, the dependency constraints may change with time. Integrating this time-dependence into our framework may further strengthen our speed limits for MPPs. 
	Additionally, we expect that incorporating finer-grained information about the topology of the dependency graph would lead to stronger SLTs. 
	However, in many cases, one does not know the full details of the dependency graph, but instead might know certain of its properties, e.g., average degree, degree distribution, or features of community structure. 
	It would therefore be valuable to extend the stochastic thermodynamics of MPPs, including the SLTs,
	to cases where one has such summary statistics of the dependency graph. 

$ $

$ $

\noindent \textbf{\large{Acknowledgements}}

\noindent	F.T. thanks David Hartich for helping translate the rate matrices in~\cite{hartich_barato_seifert_2016} to a Markov jump process in a biophysically realistic manner. This work was supported by the MIT Media Lab Consortium, Santa Fe Institute, US NSF
	Grant CHE-1648973, and FQXi Grant FQXi-RFP-IPW-1912.

	\bibliography{refs.bib}

\providecommand{\noopsort}[1]{}\providecommand{\singleletter}[1]{#1}%
\begin{thebibliography}{70}%
\makeatletter
\providecommand \@ifxundefined [1]{%
 \@ifx{#1\undefined}
}%
\providecommand \@ifnum [1]{%
 \ifnum #1\expandafter \@firstoftwo
 \else \expandafter \@secondoftwo
 \fi
}%
\providecommand \@ifx [1]{%
 \ifx #1\expandafter \@firstoftwo
 \else \expandafter \@secondoftwo
 \fi
}%
\providecommand \natexlab [1]{#1}%
\providecommand \enquote  [1]{``#1''}%
\providecommand \bibnamefont  [1]{#1}%
\providecommand \bibfnamefont [1]{#1}%
\providecommand \citenamefont [1]{#1}%
\providecommand \href@noop [0]{\@secondoftwo}%
\providecommand \href [0]{\begingroup \@sanitize@url \@href}%
\providecommand \@href[1]{\@@startlink{#1}\@@href}%
\providecommand \@@href[1]{\endgroup#1\@@endlink}%
\providecommand \@sanitize@url [0]{\catcode `\\12\catcode `\$12\catcode
  `\&12\catcode `\#12\catcode `\^12\catcode `\_12\catcode `\%12\relax}%
\providecommand \@@startlink[1]{}%
\providecommand \@@endlink[0]{}%
\providecommand \url  [0]{\begingroup\@sanitize@url \@url }%
\providecommand \@url [1]{\endgroup\@href {#1}{\urlprefix }}%
\providecommand \urlprefix  [0]{URL }%
\providecommand \Eprint [0]{\href }%
\providecommand \doibase [0]{https://doi.org/}%
\providecommand \selectlanguage [0]{\@gobble}%
\providecommand \bibinfo  [0]{\@secondoftwo}%
\providecommand \bibfield  [0]{\@secondoftwo}%
\providecommand \translation [1]{[#1]}%
\providecommand \BibitemOpen [0]{}%
\providecommand \bibitemStop [0]{}%
\providecommand \bibitemNoStop [0]{.\EOS\space}%
\providecommand \EOS [0]{\spacefactor3000\relax}%
\providecommand \BibitemShut  [1]{\csname bibitem#1\endcsname}%
\let\auto@bib@innerbib\@empty
\bibitem [{\citenamefont {Hartich}\ \emph {et~al.}(2016)\citenamefont
  {Hartich}, \citenamefont {Barato},\ and\ \citenamefont
  {Seifert}}]{hartich_barato_seifert_2016}%
  \BibitemOpen
  \bibfield  {author} {\bibinfo {author} {\bibfnamefont {D.}~\bibnamefont
  {Hartich}}, \bibinfo {author} {\bibfnamefont {A.~C.}\ \bibnamefont
  {Barato}},\ and\ \bibinfo {author} {\bibfnamefont {U.}~\bibnamefont
  {Seifert}},\ }\bibfield  {title} {\bibinfo {title} {Sensory capacity: An
  information theoretical measure of the performance of a sensor},\ }\bibfield
  {journal} {\bibinfo  {journal} {Physical Review E}\ }\textbf {\bibinfo
  {volume} {93}},\ \href {https://doi.org/10.1103/physreve.93.022116}
  {10.1103/physreve.93.022116} (\bibinfo {year} {2016})\BibitemShut {NoStop}%
\bibitem [{\citenamefont {Ashida}\ \emph {et~al.}(2021)\citenamefont {Ashida},
  \citenamefont {Aoki},\ and\ \citenamefont {Ito}}]{ashida2021experimentalSLT}%
  \BibitemOpen
  \bibfield  {author} {\bibinfo {author} {\bibfnamefont {K.}~\bibnamefont
  {Ashida}}, \bibinfo {author} {\bibfnamefont {K.}~\bibnamefont {Aoki}},\ and\
  \bibinfo {author} {\bibfnamefont {S.}~\bibnamefont {Ito}},\ }\bibfield
  {title} {\bibinfo {title} {Experimental evaluation of thermodynamic speed
  limit in living cells via information geometry},\ }\href@noop {} {\bibfield
  {journal} {\bibinfo  {journal} {bioRxiv}\ ,\ \bibinfo {pages} {2020}}
  (\bibinfo {year} {2021})}\BibitemShut {NoStop}%
\bibitem [{\citenamefont {Dutta}\ \emph {et~al.}(2020)\citenamefont {Dutta},
  \citenamefont {Sch{\"u}tz},\ and\ \citenamefont
  {Chowdhury}}]{dutta2020protein-synthesis}%
  \BibitemOpen
  \bibfield  {author} {\bibinfo {author} {\bibfnamefont {A.}~\bibnamefont
  {Dutta}}, \bibinfo {author} {\bibfnamefont {G.~M.}\ \bibnamefont
  {Sch{\"u}tz}},\ and\ \bibinfo {author} {\bibfnamefont {D.}~\bibnamefont
  {Chowdhury}},\ }\bibfield  {title} {\bibinfo {title} {Stochastic
  thermodynamics and modes of operation of a ribosome: A network theoretic
  perspective},\ }\href@noop {} {\bibfield  {journal} {\bibinfo  {journal}
  {Physical Review E}\ }\textbf {\bibinfo {volume} {101}},\ \bibinfo {pages}
  {032402} (\bibinfo {year} {2020})}\BibitemShut {NoStop}%
\bibitem [{\citenamefont {Rao}\ and\ \citenamefont
  {Esposito}(2018{\natexlab{a}})}]{rao2018chemrxn}%
  \BibitemOpen
  \bibfield  {author} {\bibinfo {author} {\bibfnamefont {R.}~\bibnamefont
  {Rao}}\ and\ \bibinfo {author} {\bibfnamefont {M.}~\bibnamefont {Esposito}},\
  }\bibfield  {title} {\bibinfo {title} {Conservation laws and work fluctuation
  relations in chemical reaction networks},\ }\href@noop {} {\bibfield
  {journal} {\bibinfo  {journal} {The Journal of chemical physics}\ }\textbf
  {\bibinfo {volume} {149}},\ \bibinfo {pages} {245101} (\bibinfo {year}
  {2018}{\natexlab{a}})}\BibitemShut {NoStop}%
\bibitem [{\citenamefont {Rao}\ and\ \citenamefont
  {Esposito}(2016)}]{rao2016chemrxnPRX}%
  \BibitemOpen
  \bibfield  {author} {\bibinfo {author} {\bibfnamefont {R.}~\bibnamefont
  {Rao}}\ and\ \bibinfo {author} {\bibfnamefont {M.}~\bibnamefont {Esposito}},\
  }\bibfield  {title} {\bibinfo {title} {Nonequilibrium thermodynamics of
  chemical reaction networks: wisdom from stochastic thermodynamics},\
  }\href@noop {} {\bibfield  {journal} {\bibinfo  {journal} {Physical Review
  X}\ }\textbf {\bibinfo {volume} {6}},\ \bibinfo {pages} {041064} (\bibinfo
  {year} {2016})}\BibitemShut {NoStop}%
\bibitem [{\citenamefont {Colegrave}(2002)}]{colegrave2002sex}%
  \BibitemOpen
  \bibfield  {author} {\bibinfo {author} {\bibfnamefont {N.}~\bibnamefont
  {Colegrave}},\ }\bibfield  {title} {\bibinfo {title} {Sex releases the speed
  limit on evolution},\ }\href@noop {} {\bibfield  {journal} {\bibinfo
  {journal} {Nature}\ }\textbf {\bibinfo {volume} {420}},\ \bibinfo {pages}
  {664} (\bibinfo {year} {2002})}\BibitemShut {NoStop}%
\bibitem [{\citenamefont {McDonald}\ \emph {et~al.}(2016)\citenamefont
  {McDonald}, \citenamefont {Rice},\ and\ \citenamefont
  {Desai}}]{mcdonald2016sex}%
  \BibitemOpen
  \bibfield  {author} {\bibinfo {author} {\bibfnamefont {M.~J.}\ \bibnamefont
  {McDonald}}, \bibinfo {author} {\bibfnamefont {D.~P.}\ \bibnamefont {Rice}},\
  and\ \bibinfo {author} {\bibfnamefont {M.~M.}\ \bibnamefont {Desai}},\
  }\bibfield  {title} {\bibinfo {title} {Sex speeds adaptation by altering the
  dynamics of molecular evolution},\ }\href@noop {} {\bibfield  {journal}
  {\bibinfo  {journal} {Nature}\ }\textbf {\bibinfo {volume} {531}},\ \bibinfo
  {pages} {233} (\bibinfo {year} {2016})}\BibitemShut {NoStop}%
\bibitem [{\citenamefont {Lloyd}(2000)}]{lloyd2000ultimate}%
  \BibitemOpen
  \bibfield  {author} {\bibinfo {author} {\bibfnamefont {S.}~\bibnamefont
  {Lloyd}},\ }\bibfield  {title} {\bibinfo {title} {Ultimate physical limits to
  computation},\ }\href@noop {} {\bibfield  {journal} {\bibinfo  {journal}
  {Nature}\ }\textbf {\bibinfo {volume} {406}},\ \bibinfo {pages} {1047}
  (\bibinfo {year} {2000})}\BibitemShut {NoStop}%
\bibitem [{\citenamefont {Markov}(2014)}]{markov2014limits}%
  \BibitemOpen
  \bibfield  {author} {\bibinfo {author} {\bibfnamefont {I.~L.}\ \bibnamefont
  {Markov}},\ }\bibfield  {title} {\bibinfo {title} {Limits on fundamental
  limits to computation},\ }\href@noop {} {\bibfield  {journal} {\bibinfo
  {journal} {Nature}\ }\textbf {\bibinfo {volume} {512}},\ \bibinfo {pages}
  {147} (\bibinfo {year} {2014})}\BibitemShut {NoStop}%
\bibitem [{\citenamefont {Laughlin}\ \emph {et~al.}(1998)\citenamefont
  {Laughlin}, \citenamefont {van Steveninck},\ and\ \citenamefont
  {Anderson}}]{laughlin1998metabolic}%
  \BibitemOpen
  \bibfield  {author} {\bibinfo {author} {\bibfnamefont {S.~B.}\ \bibnamefont
  {Laughlin}}, \bibinfo {author} {\bibfnamefont {R.~R. d.~R.}\ \bibnamefont
  {van Steveninck}},\ and\ \bibinfo {author} {\bibfnamefont {J.~C.}\
  \bibnamefont {Anderson}},\ }\bibfield  {title} {\bibinfo {title} {The
  metabolic cost of neural information},\ }\href@noop {} {\bibfield  {journal}
  {\bibinfo  {journal} {Nature neuroscience}\ }\textbf {\bibinfo {volume}
  {1}},\ \bibinfo {pages} {36} (\bibinfo {year} {1998})}\BibitemShut {NoStop}%
\bibitem [{\citenamefont {Laughlin}(2001)}]{laughlin2001energy}%
  \BibitemOpen
  \bibfield  {author} {\bibinfo {author} {\bibfnamefont {S.~B.}\ \bibnamefont
  {Laughlin}},\ }\bibfield  {title} {\bibinfo {title} {Energy as a constraint
  on the coding and processing of sensory information},\ }\href@noop {}
  {\bibfield  {journal} {\bibinfo  {journal} {Current opinion in neurobiology}\
  }\textbf {\bibinfo {volume} {11}},\ \bibinfo {pages} {475} (\bibinfo {year}
  {2001})}\BibitemShut {NoStop}%
\bibitem [{\citenamefont {Laughlin}\ and\ \citenamefont
  {Sejnowski}(2003)}]{laughlin2003communication}%
  \BibitemOpen
  \bibfield  {author} {\bibinfo {author} {\bibfnamefont {S.~B.}\ \bibnamefont
  {Laughlin}}\ and\ \bibinfo {author} {\bibfnamefont {T.~J.}\ \bibnamefont
  {Sejnowski}},\ }\bibfield  {title} {\bibinfo {title} {Communication in
  neuronal networks},\ }\href@noop {} {\bibfield  {journal} {\bibinfo
  {journal} {Science}\ }\textbf {\bibinfo {volume} {301}},\ \bibinfo {pages}
  {1870} (\bibinfo {year} {2003})}\BibitemShut {NoStop}%
\bibitem [{\citenamefont {Bassett}\ \emph {et~al.}(2011)\citenamefont
  {Bassett}, \citenamefont {Wymbs}, \citenamefont {Porter}, \citenamefont
  {Mucha}, \citenamefont {Carlson},\ and\ \citenamefont
  {Grafton}}]{bassett2011dynamic}%
  \BibitemOpen
  \bibfield  {author} {\bibinfo {author} {\bibfnamefont {D.~S.}\ \bibnamefont
  {Bassett}}, \bibinfo {author} {\bibfnamefont {N.~F.}\ \bibnamefont {Wymbs}},
  \bibinfo {author} {\bibfnamefont {M.~A.}\ \bibnamefont {Porter}}, \bibinfo
  {author} {\bibfnamefont {P.~J.}\ \bibnamefont {Mucha}}, \bibinfo {author}
  {\bibfnamefont {J.~M.}\ \bibnamefont {Carlson}},\ and\ \bibinfo {author}
  {\bibfnamefont {S.~T.}\ \bibnamefont {Grafton}},\ }\bibfield  {title}
  {\bibinfo {title} {Dynamic reconfiguration of human brain networks during
  learning},\ }\href@noop {} {\bibfield  {journal} {\bibinfo  {journal}
  {Proceedings of the National Academy of Sciences}\ }\textbf {\bibinfo
  {volume} {108}},\ \bibinfo {pages} {7641} (\bibinfo {year}
  {2011})}\BibitemShut {NoStop}%
\bibitem [{\citenamefont {Bullmore}\ and\ \citenamefont
  {Sporns}(2012)}]{bullmore2012economy}%
  \BibitemOpen
  \bibfield  {author} {\bibinfo {author} {\bibfnamefont {E.}~\bibnamefont
  {Bullmore}}\ and\ \bibinfo {author} {\bibfnamefont {O.}~\bibnamefont
  {Sporns}},\ }\bibfield  {title} {\bibinfo {title} {The economy of brain
  network organization},\ }\href@noop {} {\bibfield  {journal} {\bibinfo
  {journal} {Nature Reviews Neuroscience}\ }\textbf {\bibinfo {volume} {13}},\
  \bibinfo {pages} {336} (\bibinfo {year} {2012})}\BibitemShut {NoStop}%
\bibitem [{\citenamefont {Li}\ and\ \citenamefont
  {Dankowicz}(2019)}]{li2019opinionconsensus}%
  \BibitemOpen
  \bibfield  {author} {\bibinfo {author} {\bibfnamefont {M.}~\bibnamefont
  {Li}}\ and\ \bibinfo {author} {\bibfnamefont {H.}~\bibnamefont {Dankowicz}},\
  }\bibfield  {title} {\bibinfo {title} {Impact of temporal network structures
  on the speed of consensus formation in opinion dynamics},\ }\href@noop {}
  {\bibfield  {journal} {\bibinfo  {journal} {Physica A: Statistical Mechanics
  and its Applications}\ }\textbf {\bibinfo {volume} {523}},\ \bibinfo {pages}
  {1355} (\bibinfo {year} {2019})}\BibitemShut {NoStop}%
\bibitem [{\citenamefont {Siegenfeld}\ and\ \citenamefont
  {Bar-Yam}(2020)}]{siegenfeld2020negative}%
  \BibitemOpen
  \bibfield  {author} {\bibinfo {author} {\bibfnamefont {A.~F.}\ \bibnamefont
  {Siegenfeld}}\ and\ \bibinfo {author} {\bibfnamefont {Y.}~\bibnamefont
  {Bar-Yam}},\ }\bibfield  {title} {\bibinfo {title} {Negative representation
  and instability in democratic elections},\ }\href@noop {} {\bibfield
  {journal} {\bibinfo  {journal} {Nature Physics}\ }\textbf {\bibinfo {volume}
  {16}},\ \bibinfo {pages} {186} (\bibinfo {year} {2020})}\BibitemShut
  {NoStop}%
\bibitem [{\citenamefont {Esposito}\ and\ \citenamefont {Van~den
  Broeck}(2010{\natexlab{a}})}]{esposito2010threefaces}%
  \BibitemOpen
  \bibfield  {author} {\bibinfo {author} {\bibfnamefont {M.}~\bibnamefont
  {Esposito}}\ and\ \bibinfo {author} {\bibfnamefont {C.}~\bibnamefont {Van~den
  Broeck}},\ }\bibfield  {title} {\bibinfo {title} {Three faces of the second
  law. i. master equation formulation},\ }\href@noop {} {\bibfield  {journal}
  {\bibinfo  {journal} {Physical Review E}\ }\textbf {\bibinfo {volume} {82}},\
  \bibinfo {pages} {011143} (\bibinfo {year} {2010}{\natexlab{a}})}\BibitemShut
  {NoStop}%
\bibitem [{\citenamefont
  {Seifert}(2012)}]{seifert2012stochthermomolecularmachines}%
  \BibitemOpen
  \bibfield  {author} {\bibinfo {author} {\bibfnamefont {U.}~\bibnamefont
  {Seifert}},\ }\bibfield  {title} {\bibinfo {title} {Stochastic
  thermodynamics, fluctuation theorems and molecular machines},\ }\href@noop {}
  {\bibfield  {journal} {\bibinfo  {journal} {Reports on progress in physics}\
  }\textbf {\bibinfo {volume} {75}},\ \bibinfo {pages} {126001} (\bibinfo
  {year} {2012})}\BibitemShut {NoStop}%
\bibitem [{\citenamefont {Van~den Broeck}\ \emph {et~al.}(2013)\citenamefont
  {Van~den Broeck} \emph {et~al.}}]{vdb2013stochastic}%
  \BibitemOpen
  \bibfield  {author} {\bibinfo {author} {\bibfnamefont {C.}~\bibnamefont
  {Van~den Broeck}} \emph {et~al.},\ }\bibfield  {title} {\bibinfo {title}
  {Stochastic thermodynamics: A brief introduction},\ }\href@noop {} {\bibfield
   {journal} {\bibinfo  {journal} {Physics of Complex Colloids}\ }\textbf
  {\bibinfo {volume} {184}},\ \bibinfo {pages} {155} (\bibinfo {year}
  {2013})}\BibitemShut {NoStop}%
\bibitem [{\citenamefont {Van~den Broeck}\ and\ \citenamefont
  {Esposito}(2015)}]{vdbesposito2015ensemble}%
  \BibitemOpen
  \bibfield  {author} {\bibinfo {author} {\bibfnamefont {C.}~\bibnamefont
  {Van~den Broeck}}\ and\ \bibinfo {author} {\bibfnamefont {M.}~\bibnamefont
  {Esposito}},\ }\bibfield  {title} {\bibinfo {title} {Ensemble and trajectory
  thermodynamics: A brief introduction},\ }\href@noop {} {\bibfield  {journal}
  {\bibinfo  {journal} {Physica A: Statistical Mechanics and its Applications}\
  }\textbf {\bibinfo {volume} {418}},\ \bibinfo {pages} {6} (\bibinfo {year}
  {2015})}\BibitemShut {NoStop}%
\bibitem [{\citenamefont {Ciliberto}(2017)}]{ciliberto2017experiments}%
  \BibitemOpen
  \bibfield  {author} {\bibinfo {author} {\bibfnamefont {S.}~\bibnamefont
  {Ciliberto}},\ }\bibfield  {title} {\bibinfo {title} {Experiments in
  stochastic thermodynamics: Short history and perspectives},\ }\href@noop {}
  {\bibfield  {journal} {\bibinfo  {journal} {Physical Review X}\ }\textbf
  {\bibinfo {volume} {7}},\ \bibinfo {pages} {021051} (\bibinfo {year}
  {2017})}\BibitemShut {NoStop}%
\bibitem [{\citenamefont {Wolpert}(2019)}]{wolpert2019stofcomputation}%
  \BibitemOpen
  \bibfield  {author} {\bibinfo {author} {\bibfnamefont {D.~H.}\ \bibnamefont
  {Wolpert}},\ }\bibfield  {title} {\bibinfo {title} {The stochastic
  thermodynamics of computation},\ }\href@noop {} {\bibfield  {journal}
  {\bibinfo  {journal} {Journal of Physics A: Mathematical and Theoretical}\
  }\textbf {\bibinfo {volume} {52}},\ \bibinfo {pages} {193001} (\bibinfo
  {year} {2019})}\BibitemShut {NoStop}%
\bibitem [{\citenamefont {Horowitz}\ and\ \citenamefont
  {Gingrich}(2017)}]{horowitz2017proofFT-TUR}%
  \BibitemOpen
  \bibfield  {author} {\bibinfo {author} {\bibfnamefont {J.~M.}\ \bibnamefont
  {Horowitz}}\ and\ \bibinfo {author} {\bibfnamefont {T.~R.}\ \bibnamefont
  {Gingrich}},\ }\bibfield  {title} {\bibinfo {title} {Proof of the finite-time
  thermodynamic uncertainty relation for steady-state currents},\ }\href@noop
  {} {\bibfield  {journal} {\bibinfo  {journal} {Physical Review E}\ }\textbf
  {\bibinfo {volume} {96}},\ \bibinfo {pages} {020103} (\bibinfo {year}
  {2017})}\BibitemShut {NoStop}%
\bibitem [{\citenamefont {Horowitz}\ and\ \citenamefont
  {Gingrich}(2020)}]{horowitz2020thermodynamic}%
  \BibitemOpen
  \bibfield  {author} {\bibinfo {author} {\bibfnamefont {J.~M.}\ \bibnamefont
  {Horowitz}}\ and\ \bibinfo {author} {\bibfnamefont {T.~R.}\ \bibnamefont
  {Gingrich}},\ }\bibfield  {title} {\bibinfo {title} {Thermodynamic
  uncertainty relations constrain non-equilibrium fluctuations},\ }\href@noop
  {} {\bibfield  {journal} {\bibinfo  {journal} {Nature Physics}\ }\textbf
  {\bibinfo {volume} {16}},\ \bibinfo {pages} {15} (\bibinfo {year}
  {2020})}\BibitemShut {NoStop}%
\bibitem [{\citenamefont {Esposito}\ and\ \citenamefont {Van~den
  Broeck}(2010{\natexlab{b}})}]{esposito2010threeDFTs}%
  \BibitemOpen
  \bibfield  {author} {\bibinfo {author} {\bibfnamefont {M.}~\bibnamefont
  {Esposito}}\ and\ \bibinfo {author} {\bibfnamefont {C.}~\bibnamefont {Van~den
  Broeck}},\ }\bibfield  {title} {\bibinfo {title} {Three detailed fluctuation
  theorems},\ }\href@noop {} {\bibfield  {journal} {\bibinfo  {journal}
  {Physical review letters}\ }\textbf {\bibinfo {volume} {104}},\ \bibinfo
  {pages} {090601} (\bibinfo {year} {2010}{\natexlab{b}})}\BibitemShut
  {NoStop}%
\bibitem [{\citenamefont {Rao}\ and\ \citenamefont
  {Esposito}(2018{\natexlab{b}})}]{rao2018DFT}%
  \BibitemOpen
  \bibfield  {author} {\bibinfo {author} {\bibfnamefont {R.}~\bibnamefont
  {Rao}}\ and\ \bibinfo {author} {\bibfnamefont {M.}~\bibnamefont {Esposito}},\
  }\bibfield  {title} {\bibinfo {title} {Detailed fluctuation theorems: A
  unifying perspective},\ }\href@noop {} {\bibfield  {journal} {\bibinfo
  {journal} {Entropy}\ }\textbf {\bibinfo {volume} {20}},\ \bibinfo {pages}
  {635} (\bibinfo {year} {2018}{\natexlab{b}})}\BibitemShut {NoStop}%
\bibitem [{\citenamefont {Ito}(2018)}]{ito2018stochastic}%
  \BibitemOpen
  \bibfield  {author} {\bibinfo {author} {\bibfnamefont {S.}~\bibnamefont
  {Ito}},\ }\bibfield  {title} {\bibinfo {title} {Stochastic thermodynamic
  interpretation of information geometry},\ }\href@noop {} {\bibfield
  {journal} {\bibinfo  {journal} {Physical review letters}\ }\textbf {\bibinfo
  {volume} {121}},\ \bibinfo {pages} {030605} (\bibinfo {year}
  {2018})}\BibitemShut {NoStop}%
\bibitem [{\citenamefont {Ito}\ and\ \citenamefont
  {Dechant}(2020)}]{ito2020stochastic}%
  \BibitemOpen
  \bibfield  {author} {\bibinfo {author} {\bibfnamefont {S.}~\bibnamefont
  {Ito}}\ and\ \bibinfo {author} {\bibfnamefont {A.}~\bibnamefont {Dechant}},\
  }\bibfield  {title} {\bibinfo {title} {Stochastic time evolution, information
  geometry, and the cram{\'e}r-rao bound},\ }\href@noop {} {\bibfield
  {journal} {\bibinfo  {journal} {Physical Review X}\ }\textbf {\bibinfo
  {volume} {10}},\ \bibinfo {pages} {021056} (\bibinfo {year}
  {2020})}\BibitemShut {NoStop}%
\bibitem [{\citenamefont {Sekimoto}\ and\ \citenamefont
  {Sasa}(1997)}]{sekimoto1997complementarity}%
  \BibitemOpen
  \bibfield  {author} {\bibinfo {author} {\bibfnamefont {K.}~\bibnamefont
  {Sekimoto}}\ and\ \bibinfo {author} {\bibfnamefont {S.-i.}\ \bibnamefont
  {Sasa}},\ }\bibfield  {title} {\bibinfo {title} {Complementarity relation for
  irreversible process derived from stochastic energetics},\ }\href@noop {}
  {\bibfield  {journal} {\bibinfo  {journal} {Journal of the Physical Society
  of Japan}\ }\textbf {\bibinfo {volume} {66}},\ \bibinfo {pages} {3326}
  (\bibinfo {year} {1997})}\BibitemShut {NoStop}%
\bibitem [{\citenamefont {Aurell}\ \emph {et~al.}(2012)\citenamefont {Aurell},
  \citenamefont {Gawedzki}, \citenamefont {Mej{\'\i}a-Monasterio},
  \citenamefont {Mohayaee},\ and\ \citenamefont
  {Muratore-Ginanneschi}}]{aurell2012refined}%
  \BibitemOpen
  \bibfield  {author} {\bibinfo {author} {\bibfnamefont {E.}~\bibnamefont
  {Aurell}}, \bibinfo {author} {\bibfnamefont {K.}~\bibnamefont {Gawedzki}},
  \bibinfo {author} {\bibfnamefont {C.}~\bibnamefont {Mej{\'\i}a-Monasterio}},
  \bibinfo {author} {\bibfnamefont {R.}~\bibnamefont {Mohayaee}},\ and\
  \bibinfo {author} {\bibfnamefont {P.}~\bibnamefont {Muratore-Ginanneschi}},\
  }\bibfield  {title} {\bibinfo {title} {Refined second law of thermodynamics
  for fast random processes},\ }\href@noop {} {\bibfield  {journal} {\bibinfo
  {journal} {Journal of statistical physics}\ }\textbf {\bibinfo {volume}
  {147}},\ \bibinfo {pages} {487} (\bibinfo {year} {2012})}\BibitemShut
  {NoStop}%
\bibitem [{\citenamefont {Vo}\ \emph {et~al.}(2020)\citenamefont {Vo},
  \citenamefont {Vu},\ and\ \citenamefont {Hasegawa}}]{vo_vu_hasegawa_2020}%
  \BibitemOpen
  \bibfield  {author} {\bibinfo {author} {\bibfnamefont {V.~T.}\ \bibnamefont
  {Vo}}, \bibinfo {author} {\bibfnamefont {T.~V.}\ \bibnamefont {Vu}},\ and\
  \bibinfo {author} {\bibfnamefont {Y.}~\bibnamefont {Hasegawa}},\ }\bibfield
  {title} {\bibinfo {title} {Unified approach to classical speed limit and
  thermodynamic uncertainty relation},\ }\bibfield  {journal} {\bibinfo
  {journal} {Physical Review E}\ }\textbf {\bibinfo {volume} {102}},\ \href
  {https://doi.org/10.1103/physreve.102.062132} {10.1103/physreve.102.062132}
  (\bibinfo {year} {2020})\BibitemShut {NoStop}%
\bibitem [{\citenamefont {Van~Vu}\ and\ \citenamefont
  {Hasegawa}(2021)}]{vanvu2021geometrical}%
  \BibitemOpen
  \bibfield  {author} {\bibinfo {author} {\bibfnamefont {T.}~\bibnamefont
  {Van~Vu}}\ and\ \bibinfo {author} {\bibfnamefont {Y.}~\bibnamefont
  {Hasegawa}},\ }\bibfield  {title} {\bibinfo {title} {Geometrical bounds of
  the irreversibility in markovian systems},\ }\href@noop {} {\bibfield
  {journal} {\bibinfo  {journal} {Physical Review Letters}\ }\textbf {\bibinfo
  {volume} {126}},\ \bibinfo {pages} {010601} (\bibinfo {year}
  {2021})}\BibitemShut {NoStop}%
\bibitem [{\citenamefont {Shiraishi}\ \emph {et~al.}(2018)\citenamefont
  {Shiraishi}, \citenamefont {Funo},\ and\ \citenamefont
  {Saito}}]{shiraishi_funo_saito_2018}%
  \BibitemOpen
  \bibfield  {author} {\bibinfo {author} {\bibfnamefont {N.}~\bibnamefont
  {Shiraishi}}, \bibinfo {author} {\bibfnamefont {K.}~\bibnamefont {Funo}},\
  and\ \bibinfo {author} {\bibfnamefont {K.}~\bibnamefont {Saito}},\ }\bibfield
   {title} {\bibinfo {title} {Speed limit for classical stochastic processes},\
  }\bibfield  {journal} {\bibinfo  {journal} {Physical Review Letters}\
  }\textbf {\bibinfo {volume} {121}},\ \href
  {https://doi.org/10.1103/physrevlett.121.070601}
  {10.1103/physrevlett.121.070601} (\bibinfo {year} {2018})\BibitemShut
  {NoStop}%
\bibitem [{\citenamefont {Pietzonka}\ \emph {et~al.}(2017)\citenamefont
  {Pietzonka}, \citenamefont {Ritort},\ and\ \citenamefont
  {Seifert}}]{pietzonka2017finitetimeTUR}%
  \BibitemOpen
  \bibfield  {author} {\bibinfo {author} {\bibfnamefont {P.}~\bibnamefont
  {Pietzonka}}, \bibinfo {author} {\bibfnamefont {F.}~\bibnamefont {Ritort}},\
  and\ \bibinfo {author} {\bibfnamefont {U.}~\bibnamefont {Seifert}},\
  }\bibfield  {title} {\bibinfo {title} {Finite-time generalization of the
  thermodynamic uncertainty relation},\ }\href@noop {} {\bibfield  {journal}
  {\bibinfo  {journal} {Physical Review E}\ }\textbf {\bibinfo {volume} {96}},\
  \bibinfo {pages} {012101} (\bibinfo {year} {2017})}\BibitemShut {NoStop}%
\bibitem [{\citenamefont {Collin}\ \emph {et~al.}(2005)\citenamefont {Collin},
  \citenamefont {Ritort}, \citenamefont {Jarzynski}, \citenamefont {Smith},
  \citenamefont {Tinoco},\ and\ \citenamefont
  {Bustamante}}]{collin2005verification}%
  \BibitemOpen
  \bibfield  {author} {\bibinfo {author} {\bibfnamefont {D.}~\bibnamefont
  {Collin}}, \bibinfo {author} {\bibfnamefont {F.}~\bibnamefont {Ritort}},
  \bibinfo {author} {\bibfnamefont {C.}~\bibnamefont {Jarzynski}}, \bibinfo
  {author} {\bibfnamefont {S.~B.}\ \bibnamefont {Smith}}, \bibinfo {author}
  {\bibfnamefont {I.}~\bibnamefont {Tinoco}},\ and\ \bibinfo {author}
  {\bibfnamefont {C.}~\bibnamefont {Bustamante}},\ }\bibfield  {title}
  {\bibinfo {title} {Verification of the crooks fluctuation theorem and
  recovery of rna folding free energies},\ }\href@noop {} {\bibfield  {journal}
  {\bibinfo  {journal} {Nature}\ }\textbf {\bibinfo {volume} {437}},\ \bibinfo
  {pages} {231} (\bibinfo {year} {2005})}\BibitemShut {NoStop}%
\bibitem [{\citenamefont {Xiong}\ \emph {et~al.}(2018)\citenamefont {Xiong},
  \citenamefont {Yan}, \citenamefont {Zhou}, \citenamefont {Rehan},
  \citenamefont {Liang}, \citenamefont {Chen}, \citenamefont {Yang},
  \citenamefont {Ma}, \citenamefont {Feng},\ and\ \citenamefont
  {Vedral}}]{xiong2018experimental}%
  \BibitemOpen
  \bibfield  {author} {\bibinfo {author} {\bibfnamefont {T.}~\bibnamefont
  {Xiong}}, \bibinfo {author} {\bibfnamefont {L.}~\bibnamefont {Yan}}, \bibinfo
  {author} {\bibfnamefont {F.}~\bibnamefont {Zhou}}, \bibinfo {author}
  {\bibfnamefont {K.}~\bibnamefont {Rehan}}, \bibinfo {author} {\bibfnamefont
  {D.}~\bibnamefont {Liang}}, \bibinfo {author} {\bibfnamefont
  {L.}~\bibnamefont {Chen}}, \bibinfo {author} {\bibfnamefont {W.}~\bibnamefont
  {Yang}}, \bibinfo {author} {\bibfnamefont {Z.}~\bibnamefont {Ma}}, \bibinfo
  {author} {\bibfnamefont {M.}~\bibnamefont {Feng}},\ and\ \bibinfo {author}
  {\bibfnamefont {V.}~\bibnamefont {Vedral}},\ }\bibfield  {title} {\bibinfo
  {title} {Experimental verification of a jarzynski-related
  information-theoretic equality by a single trapped ion},\ }\href@noop {}
  {\bibfield  {journal} {\bibinfo  {journal} {Physical review letters}\
  }\textbf {\bibinfo {volume} {120}},\ \bibinfo {pages} {010601} (\bibinfo
  {year} {2018})}\BibitemShut {NoStop}%
\bibitem [{\citenamefont {Huber}\ \emph {et~al.}(2008)\citenamefont {Huber},
  \citenamefont {Schmidt-Kaler}, \citenamefont {Deffner},\ and\ \citenamefont
  {Lutz}}]{huber2008employing}%
  \BibitemOpen
  \bibfield  {author} {\bibinfo {author} {\bibfnamefont {G.}~\bibnamefont
  {Huber}}, \bibinfo {author} {\bibfnamefont {F.}~\bibnamefont
  {Schmidt-Kaler}}, \bibinfo {author} {\bibfnamefont {S.}~\bibnamefont
  {Deffner}},\ and\ \bibinfo {author} {\bibfnamefont {E.}~\bibnamefont
  {Lutz}},\ }\bibfield  {title} {\bibinfo {title} {Employing trapped cold ions
  to verify the quantum jarzynski equality},\ }\href@noop {} {\bibfield
  {journal} {\bibinfo  {journal} {Physical review letters}\ }\textbf {\bibinfo
  {volume} {101}},\ \bibinfo {pages} {070403} (\bibinfo {year}
  {2008})}\BibitemShut {NoStop}%
\bibitem [{\citenamefont {Toyabe}\ \emph {et~al.}(2010)\citenamefont {Toyabe},
  \citenamefont {Sagawa}, \citenamefont {Ueda}, \citenamefont {Muneyuki},\ and\
  \citenamefont {Sano}}]{toyabe2010experimental}%
  \BibitemOpen
  \bibfield  {author} {\bibinfo {author} {\bibfnamefont {S.}~\bibnamefont
  {Toyabe}}, \bibinfo {author} {\bibfnamefont {T.}~\bibnamefont {Sagawa}},
  \bibinfo {author} {\bibfnamefont {M.}~\bibnamefont {Ueda}}, \bibinfo {author}
  {\bibfnamefont {E.}~\bibnamefont {Muneyuki}},\ and\ \bibinfo {author}
  {\bibfnamefont {M.}~\bibnamefont {Sano}},\ }\bibfield  {title} {\bibinfo
  {title} {Experimental demonstration of information-to-energy conversion and
  validation of the generalized jarzynski equality},\ }\href@noop {} {\bibfield
   {journal} {\bibinfo  {journal} {Nature physics}\ }\textbf {\bibinfo {volume}
  {6}},\ \bibinfo {pages} {988} (\bibinfo {year} {2010})}\BibitemShut {NoStop}%
\bibitem [{\citenamefont {Barato}\ and\ \citenamefont
  {Seifert}(2015)}]{barato2015thermodynamic}%
  \BibitemOpen
  \bibfield  {author} {\bibinfo {author} {\bibfnamefont {A.~C.}\ \bibnamefont
  {Barato}}\ and\ \bibinfo {author} {\bibfnamefont {U.}~\bibnamefont
  {Seifert}},\ }\bibfield  {title} {\bibinfo {title} {Thermodynamic uncertainty
  relation for biomolecular processes},\ }\href@noop {} {\bibfield  {journal}
  {\bibinfo  {journal} {Physical review letters}\ }\textbf {\bibinfo {volume}
  {114}},\ \bibinfo {pages} {158101} (\bibinfo {year} {2015})}\BibitemShut
  {NoStop}%
\bibitem [{\citenamefont {Friedman}\ \emph {et~al.}(2020)\citenamefont
  {Friedman}, \citenamefont {Agarwalla}, \citenamefont {Shein-Lumbroso},
  \citenamefont {Tal},\ and\ \citenamefont
  {Segal}}]{friedman2020thermodynamic}%
  \BibitemOpen
  \bibfield  {author} {\bibinfo {author} {\bibfnamefont {H.~M.}\ \bibnamefont
  {Friedman}}, \bibinfo {author} {\bibfnamefont {B.~K.}\ \bibnamefont
  {Agarwalla}}, \bibinfo {author} {\bibfnamefont {O.}~\bibnamefont
  {Shein-Lumbroso}}, \bibinfo {author} {\bibfnamefont {O.}~\bibnamefont
  {Tal}},\ and\ \bibinfo {author} {\bibfnamefont {D.}~\bibnamefont {Segal}},\
  }\bibfield  {title} {\bibinfo {title} {Thermodynamic uncertainty relation in
  atomic-scale quantum conductors},\ }\href@noop {} {\bibfield  {journal}
  {\bibinfo  {journal} {Physical Review B}\ }\textbf {\bibinfo {volume}
  {101}},\ \bibinfo {pages} {195423} (\bibinfo {year} {2020})}\BibitemShut
  {NoStop}%
\bibitem [{\citenamefont {Pietzonka}\ \emph {et~al.}(2016)\citenamefont
  {Pietzonka}, \citenamefont {Barato},\ and\ \citenamefont
  {Seifert}}]{pietzonka2016universal}%
  \BibitemOpen
  \bibfield  {author} {\bibinfo {author} {\bibfnamefont {P.}~\bibnamefont
  {Pietzonka}}, \bibinfo {author} {\bibfnamefont {A.~C.}\ \bibnamefont
  {Barato}},\ and\ \bibinfo {author} {\bibfnamefont {U.}~\bibnamefont
  {Seifert}},\ }\bibfield  {title} {\bibinfo {title} {Universal bound on the
  efficiency of molecular motors},\ }\href@noop {} {\bibfield  {journal}
  {\bibinfo  {journal} {Journal of Statistical Mechanics: Theory and
  Experiment}\ }\textbf {\bibinfo {volume} {2016}},\ \bibinfo {pages} {124004}
  (\bibinfo {year} {2016})}\BibitemShut {NoStop}%
\bibitem [{\citenamefont {D'Souza}(2009)}]{dsouza2009structure}%
  \BibitemOpen
  \bibfield  {author} {\bibinfo {author} {\bibfnamefont {R.~M.}\ \bibnamefont
  {D'Souza}},\ }\bibfield  {title} {\bibinfo {title} {Structure comes to random
  graphs},\ }\href@noop {} {\bibfield  {journal} {\bibinfo  {journal} {Nature
  Physics}\ }\textbf {\bibinfo {volume} {5}},\ \bibinfo {pages} {627} (\bibinfo
  {year} {2009})}\BibitemShut {NoStop}%
\bibitem [{\citenamefont {Horowitz}\ and\ \citenamefont
  {Esposito}(2014)}]{horowitz2014bipartiteinfoflow}%
  \BibitemOpen
  \bibfield  {author} {\bibinfo {author} {\bibfnamefont {J.~M.}\ \bibnamefont
  {Horowitz}}\ and\ \bibinfo {author} {\bibfnamefont {M.}~\bibnamefont
  {Esposito}},\ }\bibfield  {title} {\bibinfo {title} {Thermodynamics with
  continuous information flow},\ }\href@noop {} {\bibfield  {journal} {\bibinfo
   {journal} {Physical Review X}\ }\textbf {\bibinfo {volume} {4}},\ \bibinfo
  {pages} {031015} (\bibinfo {year} {2014})}\BibitemShut {NoStop}%
\bibitem [{\citenamefont {Horowitz}(2015)}]{horowitz2015multipartiteinfoflow}%
  \BibitemOpen
  \bibfield  {author} {\bibinfo {author} {\bibfnamefont {J.~M.}\ \bibnamefont
  {Horowitz}},\ }\bibfield  {title} {\bibinfo {title} {Multipartite information
  flow for multiple maxwell demons},\ }\href@noop {} {\bibfield  {journal}
  {\bibinfo  {journal} {Journal of Statistical Mechanics: Theory and
  Experiment}\ }\textbf {\bibinfo {volume} {2015}},\ \bibinfo {pages} {P03006}
  (\bibinfo {year} {2015})}\BibitemShut {NoStop}%
\bibitem [{\citenamefont {Ito}\ and\ \citenamefont
  {Sagawa}(2013)}]{ito2013information}%
  \BibitemOpen
  \bibfield  {author} {\bibinfo {author} {\bibfnamefont {S.}~\bibnamefont
  {Ito}}\ and\ \bibinfo {author} {\bibfnamefont {T.}~\bibnamefont {Sagawa}},\
  }\bibfield  {title} {\bibinfo {title} {Information thermodynamics on causal
  networks},\ }\href@noop {} {\bibfield  {journal} {\bibinfo  {journal}
  {Physical Review Letters}\ }\textbf {\bibinfo {volume} {111}},\ \bibinfo
  {pages} {180603} (\bibinfo {year} {2013})}\BibitemShut {NoStop}%
\bibitem [{\citenamefont {Wolpert}(2020{\natexlab{a}})}]{wolpert_bn_2020}%
  \BibitemOpen
  \bibfield  {author} {\bibinfo {author} {\bibfnamefont {D.~H.}\ \bibnamefont
  {Wolpert}},\ }\bibfield  {title} {\bibinfo {title} {Uncertainty relations and
  fluctuation theorems for bayes nets},\ }\bibfield  {journal} {\bibinfo
  {journal} {Physical Review Letters}\ }\textbf {\bibinfo {volume} {125}},\
  \href {https://doi.org/10.1103/physrevlett.125.200602}
  {10.1103/physrevlett.125.200602} (\bibinfo {year}
  {2020}{\natexlab{a}})\BibitemShut {NoStop}%
\bibitem [{\citenamefont
  {Wolpert}(2020{\natexlab{b}})}]{wolpertstrengthenedsecondlaw}%
  \BibitemOpen
  \bibfield  {author} {\bibinfo {author} {\bibfnamefont {D.~H.}\ \bibnamefont
  {Wolpert}},\ }\bibfield  {title} {\bibinfo {title} {Strengthened landauer
  bound for composite systems},\ }\href {arXiv:2007.10950} {\bibfield
  {journal} {\bibinfo  {journal} {arXiV}\ } (\bibinfo {year}
  {2020}{\natexlab{b}})}\BibitemShut {NoStop}%
\bibitem [{\citenamefont {Wolpert}(2021)}]{wolpert_fts_mpps_2020}%
  \BibitemOpen
  \bibfield  {author} {\bibinfo {author} {\bibfnamefont {D.~H.}\ \bibnamefont
  {Wolpert}},\ }\bibfield  {title} {\bibinfo {title} {Fluctuation theorems for
  multiple systems with interdependent dynamics},\ }\href@noop {} {\bibfield
  {journal} {\bibinfo  {journal} {New Journal of Physics}\ } (\bibinfo {year}
  {2021})}\BibitemShut {NoStop}%
\bibitem [{\citenamefont {Wolpert}(2020{\natexlab{c}})}]{wolpert_min_ep_2020}%
  \BibitemOpen
  \bibfield  {author} {\bibinfo {author} {\bibfnamefont {D.~H.}\ \bibnamefont
  {Wolpert}},\ }\bibfield  {title} {\bibinfo {title} {Minimal entropy
  production rate of interacting systems},\ }\href
  {https://doi.org/10.1088/1367-2630/abc5c6} {\bibfield  {journal} {\bibinfo
  {journal} {New Journal of Physics}\ }\textbf {\bibinfo {volume} {22}},\
  \bibinfo {pages} {113013} (\bibinfo {year} {2020}{\natexlab{c}})}\BibitemShut
  {NoStop}%
\bibitem [{\citenamefont {Pilosof}\ \emph {et~al.}(2017)\citenamefont
  {Pilosof}, \citenamefont {Porter}, \citenamefont {Pascual},\ and\
  \citenamefont {K{\'e}fi}}]{pilosof2017multilayer}%
  \BibitemOpen
  \bibfield  {author} {\bibinfo {author} {\bibfnamefont {S.}~\bibnamefont
  {Pilosof}}, \bibinfo {author} {\bibfnamefont {M.~A.}\ \bibnamefont {Porter}},
  \bibinfo {author} {\bibfnamefont {M.}~\bibnamefont {Pascual}},\ and\ \bibinfo
  {author} {\bibfnamefont {S.}~\bibnamefont {K{\'e}fi}},\ }\bibfield  {title}
  {\bibinfo {title} {The multilayer nature of ecological networks},\
  }\href@noop {} {\bibfield  {journal} {\bibinfo  {journal} {Nature Ecology \&
  Evolution}\ }\textbf {\bibinfo {volume} {1}},\ \bibinfo {pages} {1} (\bibinfo
  {year} {2017})}\BibitemShut {NoStop}%
\bibitem [{\citenamefont {Delmas}\ \emph {et~al.}(2019)\citenamefont {Delmas},
  \citenamefont {Besson}, \citenamefont {Brice}, \citenamefont {Burkle},
  \citenamefont {Dalla~Riva}, \citenamefont {Fortin}, \citenamefont {Gravel},
  \citenamefont {Guimar{\~a}es~Jr}, \citenamefont {Hembry}, \citenamefont
  {Newman} \emph {et~al.}}]{delmas2019analysing}%
  \BibitemOpen
  \bibfield  {author} {\bibinfo {author} {\bibfnamefont {E.}~\bibnamefont
  {Delmas}}, \bibinfo {author} {\bibfnamefont {M.}~\bibnamefont {Besson}},
  \bibinfo {author} {\bibfnamefont {M.-H.}\ \bibnamefont {Brice}}, \bibinfo
  {author} {\bibfnamefont {L.~A.}\ \bibnamefont {Burkle}}, \bibinfo {author}
  {\bibfnamefont {G.~V.}\ \bibnamefont {Dalla~Riva}}, \bibinfo {author}
  {\bibfnamefont {M.-J.}\ \bibnamefont {Fortin}}, \bibinfo {author}
  {\bibfnamefont {D.}~\bibnamefont {Gravel}}, \bibinfo {author} {\bibfnamefont
  {P.~R.}\ \bibnamefont {Guimar{\~a}es~Jr}}, \bibinfo {author} {\bibfnamefont
  {D.~H.}\ \bibnamefont {Hembry}}, \bibinfo {author} {\bibfnamefont {E.~A.}\
  \bibnamefont {Newman}}, \emph {et~al.},\ }\bibfield  {title} {\bibinfo
  {title} {Analysing ecological networks of species interactions},\ }\href@noop
  {} {\bibfield  {journal} {\bibinfo  {journal} {Biological Reviews}\ }\textbf
  {\bibinfo {volume} {94}},\ \bibinfo {pages} {16} (\bibinfo {year}
  {2019})}\BibitemShut {NoStop}%
\bibitem [{\citenamefont {Bassett}\ and\ \citenamefont
  {Bullmore}(2006)}]{bassett2006small}%
  \BibitemOpen
  \bibfield  {author} {\bibinfo {author} {\bibfnamefont {D.~S.}\ \bibnamefont
  {Bassett}}\ and\ \bibinfo {author} {\bibfnamefont {E.}~\bibnamefont
  {Bullmore}},\ }\bibfield  {title} {\bibinfo {title} {Small-world brain
  networks},\ }\href@noop {} {\bibfield  {journal} {\bibinfo  {journal} {The
  neuroscientist}\ }\textbf {\bibinfo {volume} {12}},\ \bibinfo {pages} {512}
  (\bibinfo {year} {2006})}\BibitemShut {NoStop}%
\bibitem [{\citenamefont {Brittain}\ \emph {et~al.}(2017)\citenamefont
  {Brittain}, \citenamefont {Jones},\ and\ \citenamefont
  {Ouldridge}}]{brittain2017learningrate}%
  \BibitemOpen
  \bibfield  {author} {\bibinfo {author} {\bibfnamefont {R.~A.}\ \bibnamefont
  {Brittain}}, \bibinfo {author} {\bibfnamefont {N.~S.}\ \bibnamefont
  {Jones}},\ and\ \bibinfo {author} {\bibfnamefont {T.~E.}\ \bibnamefont
  {Ouldridge}},\ }\bibfield  {title} {\bibinfo {title} {What we learn from the
  learning rate},\ }\href@noop {} {\bibfield  {journal} {\bibinfo  {journal}
  {Journal of Statistical Mechanics: Theory and Experiment}\ }\textbf {\bibinfo
  {volume} {2017}},\ \bibinfo {pages} {063502} (\bibinfo {year}
  {2017})}\BibitemShut {NoStop}%
\bibitem [{\citenamefont {Novinger}\ \emph {et~al.}(2021)\citenamefont
  {Novinger}, \citenamefont {Suma}, \citenamefont {Sigg}, \citenamefont
  {Gonnella},\ and\ \citenamefont {Carnevale}}]{novinger2021particle}%
  \BibitemOpen
  \bibfield  {author} {\bibinfo {author} {\bibfnamefont {Q.}~\bibnamefont
  {Novinger}}, \bibinfo {author} {\bibfnamefont {A.}~\bibnamefont {Suma}},
  \bibinfo {author} {\bibfnamefont {D.}~\bibnamefont {Sigg}}, \bibinfo {author}
  {\bibfnamefont {G.}~\bibnamefont {Gonnella}},\ and\ \bibinfo {author}
  {\bibfnamefont {V.}~\bibnamefont {Carnevale}},\ }\bibfield  {title} {\bibinfo
  {title} {Particle-based ising model},\ }\href@noop {} {\bibfield  {journal}
  {\bibinfo  {journal} {Physical Review E}\ }\textbf {\bibinfo {volume}
  {103}},\ \bibinfo {pages} {012125} (\bibinfo {year} {2021})}\BibitemShut
  {NoStop}%
\bibitem [{\citenamefont {Loos}\ and\ \citenamefont
  {Klapp}(2020)}]{loos2020irreversibility}%
  \BibitemOpen
  \bibfield  {author} {\bibinfo {author} {\bibfnamefont {S.~A.}\ \bibnamefont
  {Loos}}\ and\ \bibinfo {author} {\bibfnamefont {S.~H.}\ \bibnamefont
  {Klapp}},\ }\bibfield  {title} {\bibinfo {title} {Irreversibility, heat and
  information flows induced by non-reciprocal interactions},\ }\href@noop {}
  {\bibfield  {journal} {\bibinfo  {journal} {New Journal of Physics}\ }\textbf
  {\bibinfo {volume} {22}},\ \bibinfo {pages} {123051} (\bibinfo {year}
  {2020})}\BibitemShut {NoStop}%
\bibitem [{\citenamefont {Eissfeller}\ and\ \citenamefont
  {Opper}(1994)}]{eissfeller1994asymmetricising}%
  \BibitemOpen
  \bibfield  {author} {\bibinfo {author} {\bibfnamefont {H.}~\bibnamefont
  {Eissfeller}}\ and\ \bibinfo {author} {\bibfnamefont {M.}~\bibnamefont
  {Opper}},\ }\bibfield  {title} {\bibinfo {title} {Mean-field monte carlo
  approach to the sherrington-kirkpatrick model with asymmetric couplings},\
  }\href@noop {} {\bibfield  {journal} {\bibinfo  {journal} {Physical Review
  E}\ }\textbf {\bibinfo {volume} {50}},\ \bibinfo {pages} {709} (\bibinfo
  {year} {1994})}\BibitemShut {NoStop}%
\bibitem [{\citenamefont {Sagawa}\ and\ \citenamefont
  {Ueda}(2009)}]{sagawa2009minimal}%
  \BibitemOpen
  \bibfield  {author} {\bibinfo {author} {\bibfnamefont {T.}~\bibnamefont
  {Sagawa}}\ and\ \bibinfo {author} {\bibfnamefont {M.}~\bibnamefont {Ueda}},\
  }\bibfield  {title} {\bibinfo {title} {Minimal energy cost for thermodynamic
  information processing: measurement and information erasure},\ }\href@noop {}
  {\bibfield  {journal} {\bibinfo  {journal} {Physical review letters}\
  }\textbf {\bibinfo {volume} {102}},\ \bibinfo {pages} {250602} (\bibinfo
  {year} {2009})}\BibitemShut {NoStop}%
\bibitem [{\citenamefont {Bravetti}\ and\ \citenamefont
  {Tapias}(2017)}]{bravetti2017counteradiabatic}%
  \BibitemOpen
  \bibfield  {author} {\bibinfo {author} {\bibfnamefont {A.}~\bibnamefont
  {Bravetti}}\ and\ \bibinfo {author} {\bibfnamefont {D.}~\bibnamefont
  {Tapias}},\ }\bibfield  {title} {\bibinfo {title} {Thermodynamic cost for
  classical counterdiabatic driving},\ }\href@noop {} {\bibfield  {journal}
  {\bibinfo  {journal} {Physical Review E}\ }\textbf {\bibinfo {volume} {96}},\
  \bibinfo {pages} {052107} (\bibinfo {year} {2017})}\BibitemShut {NoStop}%
\bibitem [{\citenamefont {Kolchinsky}\ and\ \citenamefont
  {Wolpert}(2020)}]{kolchinsky2020protocol-constraints}%
  \BibitemOpen
  \bibfield  {author} {\bibinfo {author} {\bibfnamefont {A.}~\bibnamefont
  {Kolchinsky}}\ and\ \bibinfo {author} {\bibfnamefont {D.~H.}\ \bibnamefont
  {Wolpert}},\ }\bibfield  {title} {\bibinfo {title} {Entropy production and
  thermodynamics of information under protocol constraints},\ }\href@noop {}
  {\bibfield  {journal} {\bibinfo  {journal} {arXiv preprint arXiv:2008.10764}\
  } (\bibinfo {year} {2020})}\BibitemShut {NoStop}%
\bibitem [{\citenamefont {Iram}\ \emph {et~al.}(2021)\citenamefont {Iram},
  \citenamefont {Dolson}, \citenamefont {Chiel}, \citenamefont {Pelesko},
  \citenamefont {Krishnan}, \citenamefont {G{\"u}ng{\"o}r}, \citenamefont
  {Kuznets-Speck}, \citenamefont {Deffner}, \citenamefont {Ilker},
  \citenamefont {Scott} \emph {et~al.}}]{iram2021controllingcounteradiabatic}%
  \BibitemOpen
  \bibfield  {author} {\bibinfo {author} {\bibfnamefont {S.}~\bibnamefont
  {Iram}}, \bibinfo {author} {\bibfnamefont {E.}~\bibnamefont {Dolson}},
  \bibinfo {author} {\bibfnamefont {J.}~\bibnamefont {Chiel}}, \bibinfo
  {author} {\bibfnamefont {J.}~\bibnamefont {Pelesko}}, \bibinfo {author}
  {\bibfnamefont {N.}~\bibnamefont {Krishnan}}, \bibinfo {author}
  {\bibfnamefont {{\"O}.}~\bibnamefont {G{\"u}ng{\"o}r}}, \bibinfo {author}
  {\bibfnamefont {B.}~\bibnamefont {Kuznets-Speck}}, \bibinfo {author}
  {\bibfnamefont {S.}~\bibnamefont {Deffner}}, \bibinfo {author} {\bibfnamefont
  {E.}~\bibnamefont {Ilker}}, \bibinfo {author} {\bibfnamefont {J.~G.}\
  \bibnamefont {Scott}}, \emph {et~al.},\ }\bibfield  {title} {\bibinfo {title}
  {Controlling the speed and trajectory of evolution with counterdiabatic
  driving},\ }\href@noop {} {\bibfield  {journal} {\bibinfo  {journal} {Nature
  Physics}\ }\textbf {\bibinfo {volume} {17}},\ \bibinfo {pages} {135}
  (\bibinfo {year} {2021})}\BibitemShut {NoStop}%
\bibitem [{\citenamefont {Diamantini}\ and\ \citenamefont
  {Trugenberger}(2014)}]{diamantini2014generalizedlandauer}%
  \BibitemOpen
  \bibfield  {author} {\bibinfo {author} {\bibfnamefont {M.~C.}\ \bibnamefont
  {Diamantini}}\ and\ \bibinfo {author} {\bibfnamefont {C.~A.}\ \bibnamefont
  {Trugenberger}},\ }\bibfield  {title} {\bibinfo {title} {Generalized landauer
  bound as a universal thermodynamic entropy in continuous phase transitions},\
  }\href@noop {} {\bibfield  {journal} {\bibinfo  {journal} {Physical Review
  E}\ }\textbf {\bibinfo {volume} {89}},\ \bibinfo {pages} {052138} (\bibinfo
  {year} {2014})}\BibitemShut {NoStop}%
\bibitem [{\citenamefont {Kempes}\ \emph {et~al.}(2017)\citenamefont {Kempes},
  \citenamefont {Wolpert}, \citenamefont {Cohen},\ and\ \citenamefont
  {P{\'e}rez-Mercader}}]{kempes2017thermodynamicefficiency}%
  \BibitemOpen
  \bibfield  {author} {\bibinfo {author} {\bibfnamefont {C.~P.}\ \bibnamefont
  {Kempes}}, \bibinfo {author} {\bibfnamefont {D.}~\bibnamefont {Wolpert}},
  \bibinfo {author} {\bibfnamefont {Z.}~\bibnamefont {Cohen}},\ and\ \bibinfo
  {author} {\bibfnamefont {J.}~\bibnamefont {P{\'e}rez-Mercader}},\ }\bibfield
  {title} {\bibinfo {title} {The thermodynamic efficiency of computations made
  in cells across the range of life},\ }\href@noop {} {\bibfield  {journal}
  {\bibinfo  {journal} {Philosophical Transactions of the Royal Society A:
  Mathematical, Physical and Engineering Sciences}\ }\textbf {\bibinfo {volume}
  {375}},\ \bibinfo {pages} {20160343} (\bibinfo {year} {2017})}\BibitemShut
  {NoStop}%
\bibitem [{\citenamefont {Wachtel}\ \emph {et~al.}(2018)\citenamefont
  {Wachtel}, \citenamefont {Rao},\ and\ \citenamefont
  {Esposito}}]{wachtel2018thermodynamically}%
  \BibitemOpen
  \bibfield  {author} {\bibinfo {author} {\bibfnamefont {A.}~\bibnamefont
  {Wachtel}}, \bibinfo {author} {\bibfnamefont {R.}~\bibnamefont {Rao}},\ and\
  \bibinfo {author} {\bibfnamefont {M.}~\bibnamefont {Esposito}},\ }\bibfield
  {title} {\bibinfo {title} {Thermodynamically consistent coarse graining of
  biocatalysts beyond michaelis--menten},\ }\href@noop {} {\bibfield  {journal}
  {\bibinfo  {journal} {New Journal of Physics}\ }\textbf {\bibinfo {volume}
  {20}},\ \bibinfo {pages} {042002} (\bibinfo {year} {2018})}\BibitemShut
  {NoStop}%
\bibitem [{\citenamefont {Yoshimura}\ and\ \citenamefont
  {Ito}(2021)}]{yoshimura2021thermodynamic}%
  \BibitemOpen
  \bibfield  {author} {\bibinfo {author} {\bibfnamefont {K.}~\bibnamefont
  {Yoshimura}}\ and\ \bibinfo {author} {\bibfnamefont {S.}~\bibnamefont
  {Ito}},\ }\bibfield  {title} {\bibinfo {title} {Thermodynamic uncertainty
  relation and thermodynamic speed limit in deterministic chemical reaction
  networks},\ }\href@noop {} {\bibfield  {journal} {\bibinfo  {journal} {arXiv
  preprint arXiv:2104.14748}\ } (\bibinfo {year} {2021})}\BibitemShut {NoStop}%
\bibitem [{\citenamefont {Gao}\ and\ \citenamefont
  {Limmer}(2021)}]{gao2021circuits}%
  \BibitemOpen
  \bibfield  {author} {\bibinfo {author} {\bibfnamefont {C.~Y.}\ \bibnamefont
  {Gao}}\ and\ \bibinfo {author} {\bibfnamefont {D.~T.}\ \bibnamefont
  {Limmer}},\ }\bibfield  {title} {\bibinfo {title} {Principles of low
  dissipation computing from a stochastic circuit model},\ }\href@noop {}
  {\bibfield  {journal} {\bibinfo  {journal} {arXiv preprint arXiv:2102.13067}\
  } (\bibinfo {year} {2021})}\BibitemShut {NoStop}%
\bibitem [{\citenamefont {Wolpert}\ and\ \citenamefont
  {Kolchinsky}(2020)}]{wolpert2020circuits}%
  \BibitemOpen
  \bibfield  {author} {\bibinfo {author} {\bibfnamefont {D.~H.}\ \bibnamefont
  {Wolpert}}\ and\ \bibinfo {author} {\bibfnamefont {A.}~\bibnamefont
  {Kolchinsky}},\ }\bibfield  {title} {\bibinfo {title} {Thermodynamics of
  computing with circuits},\ }\href@noop {} {\bibfield  {journal} {\bibinfo
  {journal} {New Journal of Physics}\ }\textbf {\bibinfo {volume} {22}},\
  \bibinfo {pages} {063047} (\bibinfo {year} {2020})}\BibitemShut {NoStop}%
\bibitem [{\citenamefont {Freitas}\ \emph {et~al.}(2020)\citenamefont
  {Freitas}, \citenamefont {Delvenne},\ and\ \citenamefont
  {Esposito}}]{freitas2020circuits}%
  \BibitemOpen
  \bibfield  {author} {\bibinfo {author} {\bibfnamefont {N.}~\bibnamefont
  {Freitas}}, \bibinfo {author} {\bibfnamefont {J.-C.}\ \bibnamefont
  {Delvenne}},\ and\ \bibinfo {author} {\bibfnamefont {M.}~\bibnamefont
  {Esposito}},\ }\bibfield  {title} {\bibinfo {title} {Stochastic
  thermodynamics of non-linear electronic circuits: A realistic framework for
  thermodynamics of computation},\ }\href@noop {} {\bibfield  {journal}
  {\bibinfo  {journal} {arXiv preprint arXiv:2008.10578}\ } (\bibinfo {year}
  {2020})}\BibitemShut {NoStop}%
\bibitem [{\citenamefont {Herpich}\ \emph {et~al.}(2020)\citenamefont
  {Herpich}, \citenamefont {Cossetto}, \citenamefont {Falasco},\ and\
  \citenamefont {Esposito}}]{herpich2020stochastic}%
  \BibitemOpen
  \bibfield  {author} {\bibinfo {author} {\bibfnamefont {T.}~\bibnamefont
  {Herpich}}, \bibinfo {author} {\bibfnamefont {T.}~\bibnamefont {Cossetto}},
  \bibinfo {author} {\bibfnamefont {G.}~\bibnamefont {Falasco}},\ and\ \bibinfo
  {author} {\bibfnamefont {M.}~\bibnamefont {Esposito}},\ }\bibfield  {title}
  {\bibinfo {title} {Stochastic thermodynamics of all-to-all interacting
  many-body systems},\ }\href@noop {} {\bibfield  {journal} {\bibinfo
  {journal} {New Journal of Physics}\ }\textbf {\bibinfo {volume} {22}},\
  \bibinfo {pages} {063005} (\bibinfo {year} {2020})}\BibitemShut {NoStop}%
\bibitem [{\citenamefont {Poulton}\ \emph {et~al.}(2019)\citenamefont
  {Poulton}, \citenamefont {Ten~Wolde},\ and\ \citenamefont
  {Ouldridge}}]{poulton2019nonequilibrium}%
  \BibitemOpen
  \bibfield  {author} {\bibinfo {author} {\bibfnamefont {J.~M.}\ \bibnamefont
  {Poulton}}, \bibinfo {author} {\bibfnamefont {P.~R.}\ \bibnamefont
  {Ten~Wolde}},\ and\ \bibinfo {author} {\bibfnamefont {T.~E.}\ \bibnamefont
  {Ouldridge}},\ }\bibfield  {title} {\bibinfo {title} {Nonequilibrium
  correlations in minimal dynamical models of polymer copying},\ }\href@noop {}
  {\bibfield  {journal} {\bibinfo  {journal} {Proceedings of the National
  Academy of Sciences}\ }\textbf {\bibinfo {volume} {116}},\ \bibinfo {pages}
  {1946} (\bibinfo {year} {2019})}\BibitemShut {NoStop}%
\bibitem [{\citenamefont {Karde{\c{s}}}\ and\ \citenamefont
  {Wolpert}(2021)}]{kardes2021turs}%
  \BibitemOpen
  \bibfield  {author} {\bibinfo {author} {\bibfnamefont {G.}~\bibnamefont
  {Karde{\c{s}}}}\ and\ \bibinfo {author} {\bibfnamefont {D.~H.}\ \bibnamefont
  {Wolpert}},\ }\bibfield  {title} {\bibinfo {title} {Thermodynamic uncertainty
  relations for multipartite processes},\ }\href@noop {} {\bibfield  {journal}
  {\bibinfo  {journal} {arXiv preprint arXiv:2101.01610}\ } (\bibinfo {year}
  {2021})}\BibitemShut {NoStop}%
\end{thebibliography}%

\end{document}